\documentclass[aps,prd,preprint,amsmath,amssymb,groupedaddress,titlepage]{revtex4-1}

\usepackage{siunitx}
\DeclareSIUnit\solarMass{\mbox{$M_\odot$}}

\usepackage{graphicx}

\usepackage{hyperref}
\hypersetup{colorlinks=true, citecolor=blue}

\def\rhoc{{\rho_\text{c}}}
\def\muc{{\check\mu}}
\def\cs2{{c_\text{s}^2}}
\def\order{{\mathcal{O}}}
\def\Mcomp{{M_\text{comp}}}

\begin{document}

% Use the \preprint command to place your local institutional report
% number in the upper righthand corner of the title page in preprint mode.
% Multiple \preprint commands are allowed.
% Use the 'preprintnumbers' class option to override journal defaults
% to display numbers if necessary
%\preprint{}

%Title of paper
\title{Tidal deformations of neutron stars with elastic crusts}

% repeat the \author .. \affiliation  etc. as needed
% \email, \thanks, \homepage, \altaffiliation all apply to the current
% author. Explanatory text should go in the []'s, actual e-mail
% address or url should go in the {}'s for \email and \homepage.
% Please use the appropriate macro foreach each type of information

% \affiliation command applies to all authors since the last
% \affiliation command. The \affiliation command should follow the
% other information
% \affiliation can be followed by \email, \homepage, \thanks as well.
\author{Fabian Gittins}
\email[]{f.w.r.gittins@soton.ac.uk}
\author{Nils Andersson}
%\homepage[]{Your web page}
%\thanks{}
%\altaffiliation{}
\affiliation{Mathematical Sciences and STAG Research Centre, 
			 University of Southampton, Southampton SO17 1BJ, United Kingdom}
\author{Jonas P. Pereira}
\affiliation{Nicolaus Copernicus Astronomical Centre, 
			 Polish Academy of Sciences, Bartycka 18, 00-716, Warsaw, Poland}

%Collaboration name if desired (requires use of superscriptaddress
%option in \documentclass). \noaffiliation is required (may also be
%used with the \author command).
%\collaboration can be followed by \email, \homepage, \thanks as well.
%\collaboration{}
%\noaffiliation

\date{\today}

\begin{abstract}
With the first detections of binary neutron star mergers by gravitational-wave 
detectors, it proves timely to consider how the internal structure of neutron 
stars affects the way in which they can be asymmetrically deformed. 
Such deformations may leave measurable imprints on gravitational-wave signals 
and can be sourced through tidal interactions or the formation of mountains. We 
detail the formalism that describes fully-relativistic neutron star models with 
elastic crusts undergoing static perturbations. This formalism primes the 
problem for studies into a variety of mechanisms that can deform a neutron star. 
We present results for a barotropic equation of state and a realistic model for 
the elastic crust, which enables us to compute relevant quantities such as the 
tidal deformability parameter. We find that the inclusion of an elastic crust 
provides a very small correction to the tidal deformability. The results allow 
us to demonstrate when and where the crust starts to fail during a binary 
inspiral and we find that the majority of the crust will remain intact up until 
merger.
\end{abstract}

% insert suggested keywords - APS authors don't need to do this
%\keywords{}

%\maketitle must follow title, authors, abstract, and keywords
\maketitle

% body of paper here - Use proper section commands
% References should be done using the \cite, \ref, and \label commands
\section{Introduction}
% Put \label in argument of \section for cross-referencing
%\section{\label{}}

The confident detections of the binary neutron star merger events GW170817 and 
GW190425 through gravitational waves has heralded in an exciting new era for 
neutron star astrophysics \cite{2017PhRvL.119p1101A, 2020ApJ...892L...3A}. Among 
other things, neutron stars provide unique probes into the equation of state of 
matter at supranuclear densities, which remains a topical issue in astrophysics 
and nuclear physics. The equation of state encodes the microscopic nuclear 
interactions and plays a fundamental role in determining the configuration of 
neutron stars. At the macroscopic level, it manifests itself through observables 
such as the mass, radius and stellar moment of inertia (see, e.g., 
\cite{2015PhRvC..91a5804S}). Terrestrial experiments are able to study and 
constrain the equation of state up to densities just below the nuclear 
saturation density $\rho_\text{sat} = \SI{2.7e14}{\gram\per\centi\metre\cubed}$ 
(see, e.g., \cite{2012PhRvC..86a5803T, 2016PrPNP..91..203B, 2016PhR...621..127L, 
2017RvMP...89a5007O} for reviews on the subject) -- markedly below the densities 
in neutron star cores. For this reason, we must look to the (neutron) stars for 
inspiration.

It has proven quite the endeavour to provide constraints on the internal 
structure of neutron stars through astrophysical observations; the observed 
range of masses is large, $M \sim \SIrange{1.1}{2.0}{\solarMass}$ 
\cite{2007PhR...442..109L}, and, problematically, it would 
seem there is no current model-independent technique to measure the radius. 
Various radius estimates have been made using X-ray spectroscopy from 
quiescent neutron stars \cite{2018MNRAS.476..421S}, thermonuclear X-ray bursts 
\cite{2010ApJ...722...33S, 2016ApJ...820...28O, 2017A&A...608A..31N} and 
accretion-powered millisecond pulsars \cite{2018A&A...618A.161S}. However, all 
these methods suffer from being susceptible to systematic errors 
\cite{2013arXiv1312.0029M, 2016EPJA...52...63M}. Promisingly, the NICER mission 
is expected to be far less prone to such biases and their first 
results have provided relatively tight bounds on the radius of PSR J0030+0451 
\cite{2019ApJ...887L..21R, 2019ApJ...887L..22R, 2019ApJ...887L..23B, 
2019ApJ...887L..24M}. Different equation-of-state 
candidates predict for a neutron star of mass $M = \SI{1.4}{\solarMass}$ a 
radius in the range of 
$\SI{9}{\kilo\metre} \lesssim R \lesssim \SI{14}{\kilo\metre}$ 
\cite{2016PhR...621..127L}.

One of the exciting prospects of gravitational-wave observations is that 
they can provide model-independent constraints on the equation of state and, 
indeed, have done in the case of GW170817 
\cite{2017PhRvL.119p1101A, 2018PhRvL.121i1102D, 2018PhRvL.121p1101A}. The 
gravitational-wave signal emitted from inspiralling neutron stars differs 
slightly from that of inspiralling black holes. The very fact that neutron stars 
are extended bodies introduces finite-size corrections to the gravitational-wave 
signal. The dominant finite-size 
effect comes from the tidal deformation that each star's gravitational field 
induces on the other. Since this effect depends on the density distribution of 
the star, it may be used as a diagnostic to probe the neutron star interior. 
However, neutron stars are believed to have solid crusts close to 
their surfaces, which introduce further complexity into prospective 
descriptions of the interior \cite{2008LRR....11...10C}. There have been a 
number of studies of perturbations of neutron stars with elastic crusts. Most of 
these have assumed the Cowling approximation 
\cite{2002A&A...395..201Y, 2017PhRvC..96f5803F} and only a few have accounted 
fully for relativistic effects \cite{1983MNRAS.203..457S, 1990MNRAS.245...82F}. 
In recent years, there have been a couple of efforts in the direction of 
understanding the impact that the inclusion of an elastic crust makes on tidal 
deformations in neutron stars \cite{2011PhRvD..84j3006P, 2019PhRvD.100d4056B}.

In this work, we detail the formalism that describes static perturbations of 
non-rotating neutron stars with an elastic component. This enables us to 
quantify the role of an elastic crust in tidal deformations. We do this for 
several different reasons. Some are technical: there are slight inconsistencies 
in the work of \cite{2011PhRvD..84j3006P}, also noted by 
\cite{2019PhRvD.100d4056B, 2019PhRvD..99b3018L}. Furthermore, we find 
discrepancies in the analysis of \cite{2019PhRvD.100d4056B}. Additionally, there 
are significant differences in results: the effects of the crust range from 
being negligible \cite{2011PhRvD..84j3006P} to being (marginally) 
detectable by third-generation gravitational-wave detectors 
\cite{2019PhRvD.100d4056B}. This is clearly an important disparity and we 
need to establish the correct answer. Intuition suggests that the result should 
be small. In fact, it is reasonably straightforward to obtain an estimate 
of the impact of the elasticity by considering \textit{f}-mode oscillations of 
an entirely elastic, incompressible star in Newtonian gravity. The result is 
then linked to the Love number (as described in \cite{2020PhRvD.101h3001A}) 
leading to a relative correction of 
$\sim 2 R \muc / (M \rho) \sim 10^{-4}$, where $\rho$ is the density 
of the star, $\muc$ is the shear modulus and we have assumed typical values 
of $M / R \sim 0.2$ and $\muc / \rho \sim 10^{-5}$. 
This estimate should be an upper limit since the entire star has been assumed to 
be elastic. This implies that this effect will be too small to be 
distinguished in observations. However, this problem is still relevant since 
this formalism provides a detailed description of the crustal structure and 
enables one to consider effects such as tidally-induced crust fractures which 
may come with associated electromagnetic signatures 
\cite{2012ApJ...749L..36P, 2012PhRvL.108a1102T}. Hence, there is a strong 
physics motivation for revisiting this problem as well.

The paper is organised as follows. To begin with, in 
Sec.~\ref{sec:Perturbations}, we describe our approach to calculating static, 
even-parity linearised perturbations of a neutron star with fluid and elastic 
layers. In Sec~\ref{sec:Tides}, we apply this formalism to calculate the tidal 
deformations of a neutron star with an elastic crust using a realistic equation 
of state and we discuss our results. We summarise and conclude in 
Sec~\ref{sec:Conclusions}.

We use the metric signature $(-, +, +, +)$ and work in geometric 
units with $G = c = 1$. We adopt the usual Einstein summation convention where 
repeated indices indicate a summation. Early Latin characters $a, b, ...$ are 
used for spacetime indices and we reserve later characters $i, j, ...$ for 
spatial indices. We use primes to denote differentiation with respect to the 
radial coordinate.

\section{\label{sec:Perturbations}Neutron star perturbations}

Since we are setting up the problem to study tidal deformations, it is 
appropriate to work in the \textit{adiabatic limit}, where the variations in the 
tidal field are assumed to be slow compared to the timescale associated with the 
star's internal response \cite{2019MNRAS.489.4043A}. For this reason, we focus 
on static perturbations of non-rotating neutron stars and we further assume that 
the unperturbed neutron star is relaxed -- that is to say, the background is 
unstrained. This is the pertinent assumption for a widely-separated binary and 
it should be valid up to the point when the crust begins to fail due to built-up 
strain. We partition our neutron stars into three layers: an inner fluid core, 
an elastic crust and an outer fluid ocean. Therefore, the equilibrium 
configuration is straightforwardly described by the usual relativistic equations 
of stellar structure: the Tolman-Oppenheimer-Volkoff equations. The elastic 
crust will manifest itself at the linear perturbation level.

\subsection{The background configuration}

The interior of a static, spherically-symmetric star is described by the 
spacetime metric $g_{a b}$ given by the line element, 
\begin{equation}
	ds^2 = g_{a b} \, dx^a dx^b = - e^\nu dt^2 + e^\lambda dr^2 
	+ r^2 (d\theta^2 + \sin^2 \theta \, d\phi^2), 
\end{equation}
where $\lambda(r)$ and $\nu(r)$ are metric functions of $r$. Since the star is 
static, the only non-vanishing component of the fluid four-velocity is the $t$ 
component. Thus, the four-velocity is 
\begin{equation}
	u^t = e^{-\nu/2}, \qquad u^i = 0.
\end{equation}
The background configuration is a perfect fluid. The stress-energy tensor for a 
perfect fluid is 
\begin{equation}
	T_{a b} = (\rho + p) u_a u_b + p \, g_{a b} 
	= \rho \, u_a u_b + p \, \bot_{a b}, 
\label{eq:StressEnergyFluid}
\end{equation}
where $\rho(r)$ is the energy density, $p(r)$ is the pressure and we have 
introduced the projection operator orthogonal to the fluid flow, 
\begin{equation}
	\bot_{a b} \equiv u_a u_b + g_{a b}, 
\end{equation}
which will be useful later.

One then solves the Einstein equations for this configuration and defines 
\begin{equation}
	e^\lambda \equiv \frac{1}{1 - 2m/r},
\end{equation}
where $m(r)$ is the gravitational mass enclosed in $r$. The mass is obtained 
from 
\begin{subequations}\label{eq:TOV}
\begin{equation}
	m' = 4 \pi r^2 \rho.
\label{eq:TOVm}
\end{equation}
The metric potential is described by 
\begin{equation}
	\nu' = \frac{2 (m + 4 \pi r^3 p)}{r (r - 2m)},
\label{eq:TOVnu}
\end{equation}
and by using the relativistic equation for hydrostatic equilibrium one finds 
\begin{equation}
	p' = - \frac{1}{2} (\rho + p) \nu' 
	= - \frac{(\rho + p) (m + 4 \pi r^3 p)}{r (r - 2m)}.
\label{eq:TOVp}
\end{equation}
\end{subequations}
Eqs.~(\ref{eq:TOV}) are the Tolman-Oppenheimer-Volkoff equations. Provided an 
equation of state one 
can solve these differential equations through numerical integration to obtain a 
description of the neutron star background. It is useful to note that 
(\ref{eq:TOVnu}) decouples from the other two equations. Therefore, one need 
only solve (\ref{eq:TOVm}) and (\ref{eq:TOVp}) to find the mass and radius 
of the star.

\subsection{\label{sec:Fluid}Fluid perturbation equations}

The standard approach to computing stellar perturbations in general relativity 
is to follow \cite{1967ApJ...149..591T}. We use the Regge-Wheeler 
gauge \cite{1957PhRv..108.1063R} and focus on static, even-parity $l \geq 2$ 
perturbations, which leads to the perturbed metric, 
\begin{equation}
	h_{a b} = 
	\begin{pmatrix}
    	e^\nu H_0	& H_1 & 0 & 0 \\
    	H_1 & e^\lambda H_2 & 0 & 0 \\
    	0   & 0 & r^2 K  & 0 \\
    	0   & 0 & 0 & r^2 \sin^2 \theta K        
	\end{pmatrix}
    Y_{l m},
\label{eq:PerturbedMetric}
\end{equation}
where $H_0(r)$, $H_1(r)$, $H_2(r)$ and $K(r)$ describe the response of the 
spacetime to the perturbations and $Y_{l m}(\theta, \phi)$ is a spherical 
harmonic. The perturbed metric is sourced by the perturbations to the 
stress-energy tensor $\delta T_a^{\hphantom{a} b}$. This coupling is described 
by the linearised Einstein equations, 
\begin{equation}
	\delta G_a^{\hphantom{a} b} = 8 \pi \, \delta T_a^{\hphantom{a} b},
\label{eq:PerturbedEinsteinEquations}
\end{equation}
where $\delta G_a^{\hphantom{a} b}$ is the perturbed Einstein tensor. The 
calculation of the perturbed Einstein tensor is rather laborious and not 
particularly insightful, so we simply state the result 
\cite{1992PhRvD..46.4289K}:
\begin{equation}
\begin{split}
	2 \delta G_a^{\hphantom{a} b} = \nabla^c \nabla_a h_c^{\hphantom{c} b} 
	&+ \nabla_c \nabla^b h_a^{\hphantom{a} c} 
	- \nabla^c \nabla_c h_a^{\hphantom{a} b} - \nabla^b \nabla_a h \\
	&- 2 R_a^{\hphantom{a} c} h_c^{\hphantom{c} b} 
	- (\nabla^d \nabla_c h_d^{\hphantom{d} c} - \nabla^c \nabla_c h 
	- R_c^{\hphantom{c} d} h_d^{\hphantom{d} c}) \, \delta_a^{\hphantom{a} b},
\label{eq:PerturbedEinstein} 
\end{split}
\end{equation}
where $h$ is the trace of the perturbed metric and $R_a^{\hphantom{a} b}$ is 
the Ricci tensor associated with the background spacetime.

To characterise the perturbations, we introduce the static displacement vector 
\cite{2011PhRvD..84j3006P}, 
\begin{equation}
	\xi^a = 
	\begin{bmatrix}
		0 \\
		r^{-1} W \\
		r^{-2} V \partial_\theta \\
		(r \sin \theta)^{-2} V \partial_\phi \\
	\end{bmatrix}
	Y_{l m}, 
\end{equation}
with functions $W(r)$ and $V(r)$ describing the radial and tangential 
displacements, respectively. In the context of relativistic perturbation theory 
the Lagrangian variation of the four-velocity is given as 
\cite{2007LRR....10....1A}
\begin{equation}
	\Delta u^a = \frac{1}{2} u^a u^b u^c \Delta g_{b c},
\end{equation}
where the Lagrangian perturbation of the metric is 
\begin{equation}
	\Delta g_{a b} = h_{a b} + 2 \nabla_{(a} \xi_{b)}.
\end{equation}
Here, the brackets around indices $(...)$ denote symmetrisation. The Lagrangian 
perturbations are straightforwardly related to the Eulerian perturbations, 
denoted with $\delta$, by $\Delta = \delta + \mathcal{L}_\xi$, where 
$\mathcal{L}_\xi$ is the Lie derivative along $\xi^a$. Thus, the Eulerian 
perturbed four-velocity is 
\begin{equation}
	\delta u^a = \bot^a_{\hphantom{a} b} \mathcal{L}_u \xi^b 
	+ \frac{1}{2} u^a u^b u^c h_{b c},
\end{equation}
which has components, 
\begin{equation}
	\delta u^t = \frac{1}{2} e^{-\nu/2} H_0 Y_{l m}, \qquad \delta u^i = 0.
\end{equation}
One should note that, given the displacement vector is static, the displacement 
vector components do not appear in the perturbed four-velocity.

A useful relation for later on comes from considering the Lagrangian change of 
the number density \cite{2007LRR....10....1A}, 
\begin{equation}
	\Delta n = - \frac{1}{2} n \bot^{a b} \Delta g_{a b}, 
\end{equation}
where $n$ is the background number density. Computing this explicitly gives 
\begin{equation}
	\Delta n = - \frac{1}{2} n \bot_\text{g} Y_{l m}, 
\label{eq:PerturbedNumberDensity}
\end{equation}
where we have defined \cite{2011PhRvD..84j3006P} 
\begin{equation}
	\bot_\text{g} \equiv \frac{2}{r^2} 
	\left[ r^2 \left( K + \frac{1}{2} H_2 \right) - l (l + 1) V + r W' 
	+ \left( 1 + \frac{1}{2} r \lambda' \right) W \right].
\label{eq:Definition}
\end{equation}
The focus of this work will be on barotropic matter, where the energy density 
is a function of only the number density, $\rho = \rho(n)$, which means that 
\begin{equation}
	\Delta \rho = \frac{d \rho}{d n} \Delta n = \mu \Delta n, 
\label{eq:Barotrope1}
\end{equation}
where $\mu$ is the chemical potential. This also leads to the relations, 
\begin{equation}
	\Delta p = \frac{d p}{d \rho} \Delta \rho = \cs2 \Delta \rho, 
	\qquad \delta p = \cs2 \delta \rho,
\label{eq:Barotrope2}
\end{equation}
where we have identified the speed of sound, $c_\text{s}$. We can use 
the Gibbs relation, $\rho + p = \mu n$, and combine 
(\ref{eq:PerturbedNumberDensity}) and (\ref{eq:Barotrope1}) to show 
\begin{equation}
	\Delta \rho = - \frac{1}{2} (\rho + p) \bot_\text{g} Y_{l m}.
\end{equation}
Therefore, by (\ref{eq:Barotrope2}) 
\begin{equation}
	\Delta p = - \frac{1}{2} (\rho + p) \cs2 \bot_\text{g} Y_{l m}.
\label{eq:Deltap1}
\end{equation}
We also have, from the relation between Lagrangian and Eulerian variations, 
\begin{equation}
	\Delta p = \delta p + \xi^r p' 
	= \delta p - \frac{1}{2 r} (\rho + p) \nu' W Y_{l m}.
\label{eq:Deltap2}
\end{equation}
We will use (\ref{eq:Deltap1}) and (\ref{eq:Deltap2}) later on to close our 
system of equations for the crustal perturbations.

The matter content of the spacetime is encoded in the stress-energy tensor. To 
complete the specification for the linearised Einstein equations we use the 
stress-energy tensor for a perfect fluid (\ref{eq:StressEnergyFluid}) to obtain 
\begin{equation}
	\delta T_a^{\hphantom{a} b} = (\delta \rho + \delta p) u_a u^b 
	+ \delta p \, \delta_a^{\hphantom{a} b} 
	+ (\rho + p) (\delta u_a u^b + u_a \delta u^b),
\label{eq:PerturbedStressEnergyFluid}
\end{equation}
where the perturbed quantities are to be expanded in spherical harmonics, e.g., 
$\delta \rho(r, \theta, \phi) \rightarrow 
\delta \rho(r) Y_{l m}(\theta, \phi)$. Note that, with such an expansion, 
a summation over all $l$, $m$ is implied. However, for this analysis, it will 
be sufficient to calculate the perturbations for a given harmonic mode.

One obtains a system of coupled ordinary differential equations by inserting 
(\ref{eq:PerturbedEinstein}) and (\ref{eq:PerturbedStressEnergyFluid}) into 
(\ref{eq:PerturbedEinsteinEquations}) which describe the perturbations in the 
fluid regions of the star. Conveniently, this system of equations simplifies to 
a single second-order differential equation 
\cite{2008ApJ...677.1216H}: 
\begin{subequations}\label{eq:FluidODEs}
\begin{equation}
\begin{split}
	H_0'' + \left( \frac{2}{r} + \frac{\nu' - \lambda'}{2} \right) H_0' 
	+ \Bigg\{ \frac{2}{r^2} - [2 + l (l + 1)] \frac{e^\lambda}{r^2}& \\
	+ \frac{9 \nu' + 5 \lambda'}{2 r} - \nu'^2 
	+ \frac{\nu' + \lambda'}{2 r \cs2} &\Bigg\} H_0 = 0.
\end{split}
\label{eq:H_0Fluid}
\end{equation}
For completeness, one can calculate the other metric perturbations from 
$H_1 = 0$, $H_2 = H_0$ and 
\begin{equation}
	[l(l + 1) - 2] e^\lambda K = r^2 \nu' H_0' 
	+ [l(l + 1) e^\lambda - 2 - r(\nu' + \lambda') + r^2 \nu'^2] H_0.
\label{eq:KFluid}
\end{equation}
\end{subequations}
As we will see later, the equations which describe the perturbations in the 
elastic crust reduce to these fluid equations. At this point, it is worth 
remarking that, given the static nature of the problem, we are unable compute 
the displacement vector in the fluid, since the functions $W$ and $V$ do not 
appear in the perturbed stress-energy tensor. This issue was somewhat confused 
in the analysis of \cite{2019PhRvD.100d4056B}, who present equations for the 
components of the displacement vector. These quantities could be calculated by 
assuming the fluid regions of the star have a small, but non-zero, shear 
modulus. We do not do this in our analysis and treat those regions as perfect 
fluids, which seems more appropriate.

\subsection{Including elasticity}

As discussed previously, the background star is assumed to be in a relaxed 
state. This means that the contribution of the 
elastic crust only enters through the perturbed stress-energy tensor. For an 
elastic material with shear modulus $\muc$, the Lagrangian perturbation of the 
anisotropic stress tensor is \cite{2019CQGra..36j5004A} 
\begin{equation}
	\Delta \pi_{a b} = - 2 \muc \Delta s_{a b},
\end{equation}
where the perturbed strain tensor $\Delta s_{a b}$ is given by 
\begin{equation}
	2 \Delta s_{a b} = \left( \bot^c_{\hphantom{c} a} \bot^d_{\hphantom{d} b} 
	- \frac{1}{3} \bot_{a b} \bot^{c d} \right) \Delta g_{c d}.
\label{eq:Strain}
\end{equation}
The anisotropic stress tensor is trace-free. Since the background is unstrained 
we simply find 
\begin{equation}
	\delta \pi_a^{\hphantom{a} b} = - \muc \left( \bot^c_{\hphantom{c} a} 
	\bot^{d b} - \frac{1}{3} \bot_a^{\hphantom{a} b} \bot^{c d} \right) 
	\Delta g_{c d}.
\label{eq:StressElastic}
\end{equation}
We note that in the equivalent expression in \cite{2011PhRvD..84j3006P} there 
is a difference of a factor of two. Summing the anisotropic stress tensor 
(\ref{eq:StressElastic}) and the fluid stress-energy tensor 
(\ref{eq:StressEnergyFluid}) and inserting these expressions into the perturbed 
Einstein equations (\ref{eq:PerturbedEinsteinEquations}) provides the 
information needed to describe perturbations in the crust.

Motivated by the analysis of \cite{1990MNRAS.245...82F}, we define the 
following dimensionless variables which are related to the radial and 
perpendicular components of the traction: 
\begin{subequations}\label{eq:TractionVariables}
\begin{align}
	T_1 Y_{l m} 
	\equiv r^2 &\delta \pi_r^{\hphantom{r} r} 
	= \frac{2 \muc}{3} [ r^2 (K - H_2) - l (l + 1) V - 2 r W' 
	+ (4 - r \lambda') W ] Y_{l m}, \\
	T_2 \partial_\theta Y_{l m} 
	\equiv r^3 &\delta \pi_r^{\hphantom{r} \theta} 
	= - \muc ( r V' - 2 V + e^\lambda W ) \partial_\theta Y_{l m}.
\end{align}
\end{subequations}
These functions, which vanish in the fluid, will help us to apply the boundary 
conditions for the problem.

Due to the introduction of the elastic crust, the perturbation equations 
become more complicated when compared to the fluid case. 
However, some of the perturbed Einstein equations remain unchanged. 
Since $\delta \pi_t^{\hphantom{t} b} = 0$, the [$tt$] component provides 
\begin{subequations}\label{eq:ElasticODEs}
\begin{equation}
\begin{split}
	e^{-\lambda} r^2 K'' 
	+ e^{-\lambda} \left( 3 - \frac{1}{2} r \lambda' \right) r K' 
	- \left[ \frac{1}{2} l (l + 1) - 1 \right]& K \\
	- e^{-\lambda} r H_2' - \left[ \frac{1}{2} l (l + 1) 
	+ e^{-\lambda} (1 - r\lambda') \right]& H_2 = - 8 \pi r^2 \delta\rho.
\end{split}
\label{eq:PEEtt}
\end{equation}
Because $\delta \pi_a^{\hphantom{a} b}$ is traceless, we can take the trace of 
the perturbed Einstein equations to obtain another equation that has no explicit 
dependence on the elasticity. We combine the trace with (\ref{eq:PEEtt}) to 
obtain 
\begin{equation}
\begin{split}
	- r^2 H_0'' 
	+ \left[ r \left( \frac{1}{2} \lambda' - \nu' \right) - 2 \right] r H_0' 
	+ l (l + 1) e^\lambda &H_0 \\
	- \frac{1}{2} r^2 \nu' H_2' + [2 (e^\lambda - 1) 
	- r (3\nu' + \lambda')] H_2 + r^2 \nu' &K' 
	= 8 \pi r^2 e^\lambda (\delta\rho + 3 \delta p).
\end{split}
\label{eq:PEEtrace}
\end{equation}
Furthermore, we find from the [$tr$] component that, as in the fluid case, 
$H_1$ vanishes.

Now, we consider the non-zero components of $\delta \pi_a^{\hphantom{a} b}$. 
The difference between the [$\theta\theta$] and [$\phi\phi$] components leads 
to the algebraic relation, 
\begin{equation}
    H_2 - H_0 = 32 \pi \muc V,
\label{eq:PEEdifference}
\end{equation}
which will be useful to eliminate $H_2$ from our equations. We can use the 
[$r\theta$] component and (\ref{eq:PEEdifference}) to provide 
\begin{equation}
	K' = H_0' + \nu' H_0 + \frac{16 \pi}{r} (2 + r\nu') \muc V 
	- \frac{16 \pi}{r} T_2.
\label{eq:PEErthetaa}
\end{equation}
The sum of the [$\theta\theta$] and [$\phi\phi$] components gives 
\begin{equation}
\begin{split}
	\delta p = \frac{e^{-\lambda} (\nu' + \lambda')}{16 \pi r} H_0 
	+ \frac{e^{-\lambda}}{r^2} \bigg\{ e^\lambda [2 - l (l + 1)] 
	\muc V& \\
	+ \frac{e^\lambda}{2} T_1 - r T_2' 
	- \left[ \frac{1}{2} r (\nu' - \lambda') + 1 \right] T_2& \bigg\},
\end{split}
\label{eq:PEEsum}
\end{equation}
where we have simplified using (\ref{eq:PEEdifference}) and 
(\ref{eq:PEErthetaa}). The final equation we will use from the perturbed 
Einstein equations is the [$rr$] component combined with 
Eqs.~(\ref{eq:PEEdifference})--(\ref{eq:PEEsum}), 
\begin{equation}
\begin{split}
	[l (l + 1) - 2] e^\lambda K = r^2 \nu' &H_0' 
	+ [l (l + 1) e^\lambda - 2 - r (\nu' + \lambda') + r^2 \nu'^2] H_0 \\ 
	+ 16 &\pi \{ [l (l + 1) - 2] e^\lambda + r^2 \nu'^2 \} \muc V \\ 
	- 24 &\pi e^\lambda T_1 + 16 \pi r T_2' 
	- 8 \pi [2 + r (\nu' + \lambda')] T_2.
\end{split}
\label{eq:PEErr}
\end{equation}
\end{subequations}
When $\muc = 0$, this reduces to (\ref{eq:KFluid}), as expected. Note that the 
corresponding equation in \cite{2011PhRvD..84j3006P} differs from this by 
missing a factor of $e^\lambda$ in the coefficients of $T_1$ and $V$.

The next step is to formulate the system of equations in a way that is 
straightforward to integrate numerically. Clearly, there is a lot of freedom 
in how one can do this. We choose to work with the functions 
($H_0'$, $H_0$, $K$, $W$, $V$, $T_2$) as our integration variables. It is 
useful to observe that through (\ref{eq:PEEdifference}) one can reduce the 
order of the system to eliminate $H_2$. In contrast to the fluid case, we are 
able to solve for the components of the displacement vector by using the 
definitions of the traction variables (\ref{eq:TractionVariables}). To be 
precise, we can integrate 
\begin{subequations}\label{eq:FinalODEs}
\begin{equation}
	W' - \left( \frac{2}{r} - \frac{\lambda'}{2} \right) W 
	= \frac{1}{2} r (K - H_0) - \left[ 16 \pi r \muc 
	+ \frac{l (l + 1)}{2 r} \right] V - \frac{3}{4 \muc r} T_1 
\end{equation}
and 
\begin{equation}
	V' - \frac{2}{r} V = - \frac{e^\lambda}{r} W - \frac{1}{\muc r} T_2.
\end{equation}
We obtain an algebraic relation by combining (\ref{eq:PEEsum}) and 
(\ref{eq:PEErr}) in such a way as to remove $T_2'$. This gives us an equation 
which involves $\delta p$ and $T_1$, 
\begin{equation}
\begin{split}
	16 \pi r^2 e^\lambda \delta p = r^2 \nu' &H_0' 
	+ [l (l + 1) e^\lambda - 2 + r^2 \nu'^2] H_0 
	+ [2 - l (l + 1)] e^\lambda K \\
	+ 16 &\pi r^2 \nu'^2 \muc V - 16 \pi e^\lambda T_1 
	- 16 \pi (2 + r \nu') T_2.
\end{split}
\label{eq:AlgebraicRelationa}
\end{equation}
From (\ref{eq:PEEsum}) we can obtain an equation to integrate for $T_2$, 
\begin{equation}
	T_2' + \left( \frac{\nu' - \lambda'}{2} + \frac{1}{r} \right) T_2 
	= - r e^\lambda \delta p + \frac{\nu' + \lambda'}{16 \pi} H_0 
	+ \frac{e^\lambda}{r} [2 - l (l + 1)] \muc V 
	+ \frac{e^\lambda}{2 r} T_1.
\label{eq:T_2Elastic}
\end{equation}
We combine Eqs.~(\ref{eq:PEEtrace})--(\ref{eq:PEErthetaa}) to get 
\begin{equation}
\begin{split}
	H_0''& + \left( \frac{2}{r} + \frac{\nu' - \lambda'}{2} \right) H_0' 
	+ \Bigg\{ \frac{2}{r^2} - [2 + l (l + 1)] \frac{e^\lambda}{r^2} 
	+ \frac{3 \nu' + \lambda'}{r} - \nu'^2 \Bigg\} H_0 \\
	= &- 8 \pi \left[ e^\lambda \left( 3 
	+ \frac{1}{\cs2} \right) \delta p + 2 \nu' (\muc V)' 
	+ 8 \left( \frac{1 - e^\lambda}{r^2} + \frac{2 \nu' + \lambda'}{2 r} 
	- \frac{1}{4} \nu'^2 \right) \muc V + \frac{2 \nu'}{r} T_2 \right].
\end{split}
\label{eq:H_0Elastic}
\end{equation}
In the fluid, where the shear modulus vanishes, one can verify that 
(\ref{eq:T_2Elastic}) and (\ref{eq:H_0Elastic}) reduce to give 
(\ref{eq:H_0Fluid}). The final equation we need from the perturbed Einstein 
equations (\ref{eq:PEErthetaa}), usefully, needs no further alteration, 
\begin{equation}
	K' = H_0' + \nu' H_0 + \frac{16 \pi}{r} (2 + r\nu') \muc V 
	- \frac{16 \pi}{r} T_2.
\label{eq:PEErthetab}
\end{equation}
To close this system of equations we need to consider the thermodynamics. We 
can use (\ref{eq:Deltap1}) and (\ref{eq:Deltap2}) to obtain a second algebraic 
relation involving $\delta p$ and $T_1$ by substituting for $W'$ in 
$\bot_\text{g}$ (\ref{eq:Definition}), 
\begin{equation}
	\frac{3}{4 \muc} T_1 
	= \frac{r^2}{(\rho + p) \cs2} \delta p 
	+ \frac{3}{2} r^2 K - \frac{3}{2} l (l + 1) V 
	+ \left(3 - \frac{r \nu'}{2 \cs2}\right) W.
	\label{eq:AlgebraicRelationb}
\end{equation}
\end{subequations}
We use (\ref{eq:AlgebraicRelationa}) and 
(\ref{eq:AlgebraicRelationb}) to determine $\delta p$ and $T_1$.
Eqs.~(\ref{eq:FinalODEs}) fully specify the elastic 
perturbation problem.

\subsection{Boundary conditions}

To solve for the perturbations throughout the star one needs to solve 
Eqs.~(\ref{eq:FluidODEs}) in the fluid regions and Eqs.~(\ref{eq:FinalODEs}) in 
the crust. At the centre of the star, the equations are singular and so we 
demand regularity to obtain the initial condition, for small $r$, 
\begin{equation}
	H_0(r) = a_0 r^l [ 1 + \order(r^2) ], 
\label{eq:InitialCondition}
\end{equation}
where $a_0$ is a constant. This can be derived by considering a power-series 
expansion for small $r$. We note that, \textit{a priori}, we do not know 
the amplitude of the perturbations since we have not specified the mechanism 
that sources them. To single out a unique solution from this one-parameter 
family of solutions one must match the interior solution to the exterior vacuum 
solution, which, in the case of tidal deformations, is sourced by the tidal 
potential of the companion star \cite{2008ApJ...677.1216H}.

There are two fluid-elastic interfaces in the neutron star, where one has to 
consider the continuity of the perturbed variables. From our assumption that 
the background star is in a relaxed state, we know that the background 
quantities will be continuous across an interface. Of course, should one use an 
equation of state that involves discontinuities at such an interface, that 
would need to be taken into account. We do not consider such possibilities here, 
but do so in \cite{2020arXiv200310781P}.

In order to determine how the perturbed quantities behave at an interface, we 
must calculate the first and second fundamental forms and demand that they are 
continuous across the interface. We describe this calculation in detail in 
Appendix~\ref{app:Interface}. The first fundamental form implies that the 
functions $H_0$, $K$ and $W$ are continuous. From the second fundamental form 
we obtain continuity of the radial, $(T_1 + r^2 \delta p)$, and tangential 
traction, $T_2$. We will assume that the shear modulus is non-zero throughout 
the crust and, therefore, must be discontinuous at a fluid-elastic boundary. 
Alternatively, one could consider a shear modulus that smoothly goes to zero at 
an interface. In this case, one might assume that the traction conditions would 
be trivially satisfied. This may be more realistic, but it is difficult to model 
as we do not have a description for the precise core-crust transition.

In the core we calculate $H_0'$, $H_0$ and $K$, but in the crust the order of 
the system increases as we need to calculate the additional functions $W$, $V$ 
and $T_2$. We know that in the fluid the shear modulus vanishes and so $T_2 = 0$ 
at both the core-crust and crust-ocean interfaces. We can use 
(\ref{eq:H_0primeb}) along with (\ref{eq:KFluid}) to obtain an expression which 
is true in the elastic crust at an interface, 
\begin{equation}
	r^2 \nu' H_0' = - [ l (l + 1) e^\lambda - 2 - r (\nu' + \lambda') 
	+ r^2 \nu'^2 ] H_0 + [ l (l + 1) - 2 ] e^\lambda K 
	- 16 \pi r^2 \nu'^2 \muc V.
\label{eq:H_0primeCondition}
\end{equation}
With the six boundary conditions -- continuity of $H_0$ and $K$ at the 
core-crust interface, and the constraints at both interfaces: $T_2 = 0$ and 
(\ref{eq:H_0primeCondition}) -- the system is well posed as a boundary-value 
problem. 

The surface of the perturbed configuration is defined to be where the Lagrangian 
variation of the pressure vanishes, $\Delta p = 0$. Because of 
(\ref{eq:Deltap1}), this conveniently coincides with the definition of the 
surface for the background star, $p = 0$. We describe our numerical approach to 
solving this problem in detail in Appendix~\ref{app:Numerical}.

\section{\label{sec:Tides}Tidal deformations}

The formalism we have detailed in Sec.~\ref{sec:Perturbations} can be applied 
to a variety of problems, such as tidal deformations and mountains on neutron 
stars. In this work, we specialise the perturbations to those sourced by 
tides in binary systems.

\subsection{\label{sec:k_2}The tidal deformability}

A star of mass $M$ and radius $R$ in a time-independent, external tidal field 
$\mathcal{E}_{i j}$ will 
develop a quadrupole moment $Q_{i j}$ in response. To linear order, one can 
relate the quadrupole moment to the tidal field by \cite{2008ApJ...677.1216H}
\begin{equation}
	Q_{i j} = - \frac{2}{3} k_2 R^5 \mathcal{E}_{i j}, 
\end{equation}
where we have introduced the tidal Love number $k_2$. We briefly review 
the procedure of calculating the tidal Love number below, closely following the 
explanation in \cite{2008ApJ...677.1216H}. For other detailed 
discussions on the subject we refer the reader to 
\cite{2009PhRvD..80h4018B, 2009PhRvD..80h4035D}.

The Love number can be extracted from the asymptotic behaviour of the metric. In 
asymptotically Cartesian and mass-centred coordinates, one can write 
\cite{1980RvMP...52..299T}
\begin{equation}
    - \frac{1 + \textsl{g}_{t t}}{2} = - \frac{M}{r} - \frac{3 Q_{i j}}{2 r^3}
    \left( n^i n^j - \frac{1}{3} \delta^{i j} \right) + \order(1 / r^4)
    + \frac{1}{2} r^2 \mathcal{E}_{i j} n^i n^j + \order(r^3),
\label{eq:AsymptoticRestFrame}
\end{equation}
where $x^i$ is the vector that points from the origin to $r$, $n^i = x^i / r$ is 
the corresponding unit vector and $\textsl{g}_{a b} = g_{a b} + h_{a b}$ 
corresponds to the full metric up to first order. In the vacuum exterior, one 
should note that $\nu = - \lambda$, therefore, (\ref{eq:H_0Fluid}) reduces to 
\begin{equation}
    H_0'' + \left( \frac{2}{r} - \lambda' \right) H_0'
    - \left[ l (l + 1) \frac{e^\lambda}{r^2} + \lambda'^2 \right] H_0 = 0.
\label{eq:H_0ExteriorODE}
\end{equation}
The solution to (\ref{eq:H_0ExteriorODE}) may be expressed in terms of the 
associated Legendre polynomials $\mathcal{Q}_{\alpha \beta}(r / M - 1)$ and 
$\mathcal{P}_{\alpha \beta}(r / M - 1)$ with $\alpha = l$, $\beta = 2$, 
\begin{equation}
	H_0(r) = c_1 \mathcal{Q}_{l 2}(r / M - 1) 
	+ c_2 \mathcal{P}_{l 2}(r / M - 1), 
\end{equation}
which gives, when we specialise to quadrupolar ($l = 2$) perturbations, 
\begin{equation}
\begin{split}
		H_0(r) = c_1 \left( \frac{r}{M} \right)^2 \left( 1 - \frac{2 M}{r} \right) 
		\Bigg[ &- \frac{M (M - r) (2 M^2 + 6 M r - 3 r^2)}{r^2 (2 M - r)^2} \\
		&+ \frac{3}{2} \ln \left( \frac{r}{r - 2 M} \right) \Bigg] 
		+ 3 c_2 \left( \frac{r}{M} \right)^2 \left( 1 - \frac{2 M}{r} \right), 
\end{split}
\label{eq:H_0Exterior}
\end{equation}
where $c_1$ and $c_2$ are constants to be determined. The asymptotic 
behaviour of (\ref{eq:H_0Exterior}) is 
\begin{equation}
	H_0(r) = \frac{8}{5} \left( \frac{M}{r} \right)^3 c_1 
	+ \order[(M / r)^4] + 3 \left( \frac{r}{M} \right)^2 c_2 
	+ \order(r / M).
\label{eq:H_0Asymptotic}
\end{equation}
One can decompose the tensor multipole moments as 
\begin{subequations}\label{eq:Decomposition}
\begin{gather}
	\mathcal{E}_{i j} 
	= \sum_{m = -2}^2 \mathcal{E}_{2 m} \mathcal{Y}_{i j}^{2 m}, \\
	Q_{i j} = \sum_{m = -2}^2 Q_{2 m} \mathcal{Y}_{i j}^{2 m}, 
\end{gather}
\end{subequations}
where the symmetric, trace-free tensors $\mathcal{Y}_{i j}^{l m}$ are defined 
by \cite{1980RvMP...52..299T}
\begin{equation}
	Y_{l m} = \mathcal{Y}_{i j}^{l m} n^i n^j.
\end{equation}
One is free to assume that only one $\mathcal{E}_{2 m}$ is non-vanishing, 
without any loss of generality. Making use of the decomposition of the multipole 
moments (\ref{eq:Decomposition}), one can insert (\ref{eq:H_0Asymptotic}) into 
(\ref{eq:AsymptoticRestFrame}) to show 
\begin{subequations}\label{eq:Coefficients}
\begin{gather}
	c_1 = \frac{15}{8} \frac{1}{M^3} Q_{2 m}, \\
	c_2 = - \frac{1}{3} M^2 \mathcal{E}_{2 m},\label{eq:c_2a}
\end{gather}
\end{subequations}
and, thus, obtain 
\begin{equation}
	\frac{c_1}{c_2} = \frac{15}{4} \frac{k_2}{C^5}, 
\label{eq:Ratio}
\end{equation}
where $C \equiv M / R$ is the star's compactness. Because $H_0$ and $H_0'$ 
are continuous between the interior and the vacuum at the surface, we can use 
(\ref{eq:H_0Exterior}) to determine the ratio $c_1 / c_2$ in 
terms of the interior solutions at $r = R$. This gives the result 
\cite{2008ApJ...677.1216H}
\begin{equation}
\begin{split}
	k_2 = \frac{8 C^5}{5} (1& - 2 C)^2 [2 + 2 C (y - 1) - y] 
	\Big\{ 2 C [6 - 3 y + 3 C (5 y - 8)] \\
	&+ 4 C^3 [13 - 11 y + C (3 y - 2) + 2 C^2 (1 + y)] \\
	&+ 3 (1 - 2 C)^2 [2 - y + 2 C (y - 1)] \ln(1 - 2 C) \Big\}^{-1},
\end{split}
\label{eq:k_2}
\end{equation}
where we have introduced the parameter $y \equiv R H_0'(R) / H_0(R)$. It is 
interesting to note that for the computation of the Love number the 
amplitude $a_0$ in the initial condition (\ref{eq:InitialCondition}) may be 
chosen freely. The reason for this is intuitive. Since the tidal Love number is 
a measure of how deformable a star is in the presence of a quadrupolar field, it 
is independent of the exact details of an external field and, therefore, the 
calculation of this quantity is insensitive to the magnitude. We see this in 
(\ref{eq:k_2}) as the ratio $y$ means that dependence on $a_0$ exactly cancels. 
For our analysis, we will focus on the dimensionless tidal deformability 
parameter, 
\begin{equation}
	\Lambda = \frac{2}{3} \frac{k_2}{C^5},
\end{equation}
to enable direct comparison with gravitational-wave constraints 
(see, e.g., \cite{2018PhRvL.121p1101A}).

Note that the expression (\ref{eq:k_2}) is derived under the assumption that 
$H_0'$ is continuous across the surface. This is contingent on the final 
stellar layer having a vanishing shear modulus, or equivalently being a fluid 
(\ref{eq:H_0primeb}). This detail was overlooked by \cite{2019PhRvD.100d4056B}, 
who treat the outer layer to be the elastic crust and yet use (\ref{eq:k_2}) to 
calculate the tidal Love number. One could of course compute the Love number for 
a star with an elastic outer region; however, one would need to incorporate the 
discontinuity of $H_0'$ using (\ref{eq:H_0primeb}) by taking into account the 
value of $V$ at the surface.

To accurately prescribe the crust-ocean transition, we consider the melting 
point of the crust. The Coulomb lattice melts when the thermal energy, 
\begin{equation}
	E_\text{th} = k_\text{B} T,
\end{equation}
where $k_\text{B}$ is Boltzmann's constant and $T$ is the temperature, exceeds 
the interaction energy of the lattice, 
\begin{equation}
	E_\text{Coul} = \frac{1}{4 \pi \varepsilon_0} \frac{Z^2 e^2}{a}, 
\end{equation}
where $Z$ is the proton number, $e$ is the unit charge, $a$ is the mean 
spacing between nuclei and $\varepsilon_0$ is the permittivity of free space, 
by a critical factor $1/\Gamma$, 
\begin{equation}
	E_\text{th} \geq \frac{1}{\Gamma} E_\text{Coul}, 
\end{equation}
where $\Gamma \approx 173$. We assume that the crust forms a body-centred cubic 
lattice, which has two nuclei per unit cube, so given the number density of 
nuclei, $n_\text{N}$, we have 
\begin{equation}
	n_\text{N} a^3 = 2.
\end{equation}
The density at which the crust begins to melt is, therefore, obtained from 
\begin{equation}
	\rho_\text{top} = A m_\text{u} n_\text{N} 
	= 2 A m_\text{u} \left( 4 \pi \varepsilon_0 
		\frac{\Gamma k_\text{B} T}{Z^2 e^2} \right)^3 
	\approx \num{6.72e5} \ \left( \frac{A}{56} \right) 
		\left( \frac{Z}{26} \right)^{-2/3} 
		\left( \frac{T}{\SI{e7}{\kelvin}} \right)^3 
		\ \si{\gram\per\centi\metre\cubed}, 
\end{equation}
where $A$ is the nucleon number and $m_\text{u}$ is the atomic mass unit. For 
our prescription, we assume that the outer parts of the crust are composed of 
iron, $Z = 26$ and $A = 56$, and a temperature of $T = \SI{e7}{\kelvin}$.

\begin{figure}
	\includegraphics[width=0.7\textwidth]{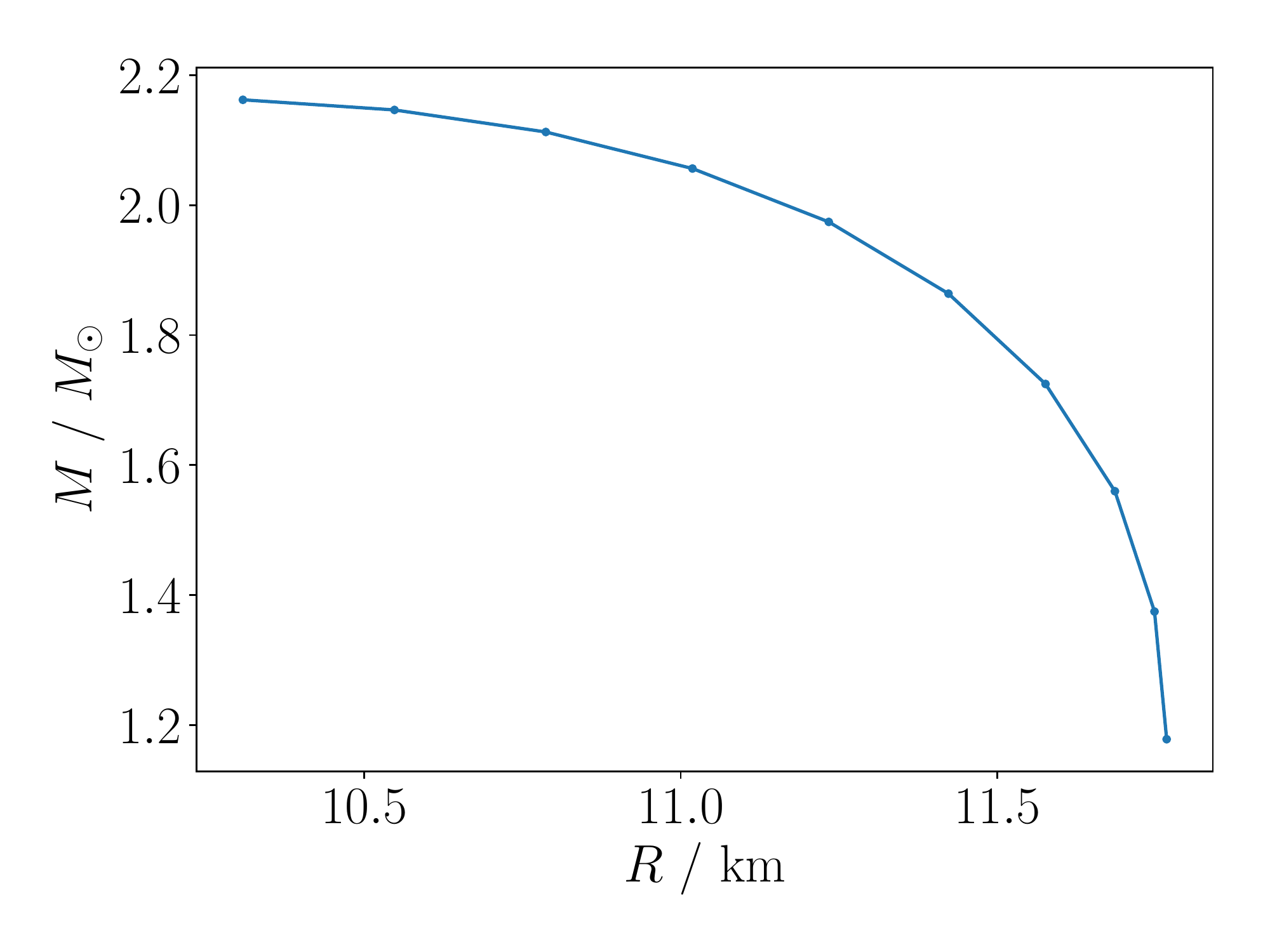}%
	\caption{\label{fig:MassRadius}The mass-radius diagram for the central 
			 densities considered, showing that the stellar models considered 
			 are stable to radial perturbations.}
\end{figure}

The star is defined to have: (i) a fluid core for 
$\rhoc \geq \rho > \rho_\text{base}$, (ii) an elastic crust in the region 
$\rho_\text{base} \geq \rho > \rho_\text{top}$ and (iii) a fluid ocean for 
$\rho \leq \rho_\text{top}$, where the base of the crust is defined to be 
$\rho_\text{base} = \SI{1.3e14}{\gram\per\centi\metre\cubed}$. We use the 
BSk20 analytic equation of state \cite{2013A&A...560A..48P} for the high-density 
fluid core and the equation-of-state table from \cite{2001A&A...380..151D} for 
the low-density regions. We parametrise each stellar model according to its 
central density and integrate Eqs.~(\ref{eq:TOV}) for the background. The 
background is solved along with Eqs.~(\ref{eq:FluidODEs}) in the fluid 
regions of the star and Eqs.~(\ref{eq:FinalODEs}) in the crust. The results of 
the integrations are summarised in Table~\ref{tab:Results}. The mass and 
radius of each stellar model is presented in Fig.~\ref{fig:MassRadius} to show 
that they are all stable to radial perturbations.

\begin{table}%[H] add [H] placement to break table across pages
	\caption{\label{tab:Results}Results of the numerical integrations of 
	the perturbation equations using the BSk20 equation of state for the 
	core \cite{2013A&A...560A..48P} and the equation of state from 
	\cite{2001A&A...380..151D} for the low-density layers of the star. Each 
	stellar model is determined by the central 
	density $\rhoc$. We provide the radius $R$, mass $M$, compactness $C$ and 
	crustal thickness $\Delta R_\text{c}$ for each star. The tidal deformability 
	for the fluid stars, $\Lambda_\text{fluid}$, and those with elastic crusts, 
	$\Lambda_\text{crust}$, are shown, along with the relative difference 
	between them, where 
	$\Delta\Lambda \equiv \Lambda_\text{crust} - \Lambda_\text{fluid}$. From the 
	differences between the tidal deformabilities, we see that the correction 
	due to the presence of a crust is very small.}
	{\footnotesize
\begin{ruledtabular}
\begin{tabular}{ c c c c r r c c }
	$\rhoc$ / \SI{e15}{\gram\per\centi\metre\cubed} & $R$ / \si{\kilo\metre} & 
	$M$ / \si{\solarMass} & $C$ & 
	\multicolumn{1}{c}{$\Lambda_\text{crust}$} & 
	\multicolumn{1}{c}{$\Lambda_\text{fluid}$} & 
	$\Delta\Lambda / \Lambda_\text{fluid}$ & 
	$\Delta R_\text{c}$ / \si{\kilo\metre} \\
	\hline 
	\num{2.500} & \num{10.309} & \num{2.162} & \num{0.310} & 
	\num{3.95452368861} & \num{3.95452375613} & \num{-1.707e-08} & \num{0.278} \\
	\num{2.203} & \num{10.548} & \num{2.146} & \num{0.301} & 
	\num{5.47304743630} & \num{5.47304753879} & \num{-1.873e-08} & \num{0.307} \\
	\num{1.941} & \num{10.787} & \num{2.112} & \num{0.289} & 
	\num{8.01552675384} & \num{8.01552692367} & \num{-2.119e-08} & \num{0.343} \\
	\num{1.710} & \num{11.019} & \num{2.056} & \num{0.276} & 
	\num{12.49881579778} & \num{12.49881610913} & \num{-2.491e-08} & \num{0.391} \\
	\num{1.507} & \num{11.234} & \num{1.974} & \num{0.260} & 
	\num{20.87848456088} & \num{20.87848520087} & \num{-3.065e-08} & \num{0.451} \\
	\num{1.327} & \num{11.423} & \num{1.864} & \num{0.241} & 
	\num{37.57928548491} & \num{37.57928697864} & \num{-3.975e-08} & \num{0.529} \\
	\num{1.170} & \num{11.576} & \num{1.725} & \num{0.220} & 
	\num{73.25641923536} & \num{73.25642323791} & \num{-5.464e-08} & \num{0.629} \\
	\num{1.031} & \num{11.686} & \num{1.560} & \num{0.197} & 
	\num{155.28383741339} & \num{155.28384983242} & \num{-7.998e-08} & \num{0.758} \\
	\num{0.908} & \num{11.748} & \num{1.375} & \num{0.173} & 
	\num{358.77773415782} & \num{358.77777902699} & \num{-1.251e-07} & \num{0.926} \\
	\num{0.800} & \num{11.768} & \num{1.178} & \num{0.148} & 
	\num{903.80359991409} & \num{903.80378919034} & \num{-2.094e-07} & \num{1.144} \\
\end{tabular}
\end{ruledtabular}
	}
\end{table}

\begin{figure}
	\includegraphics[width=0.49\textwidth, keepaspectratio]{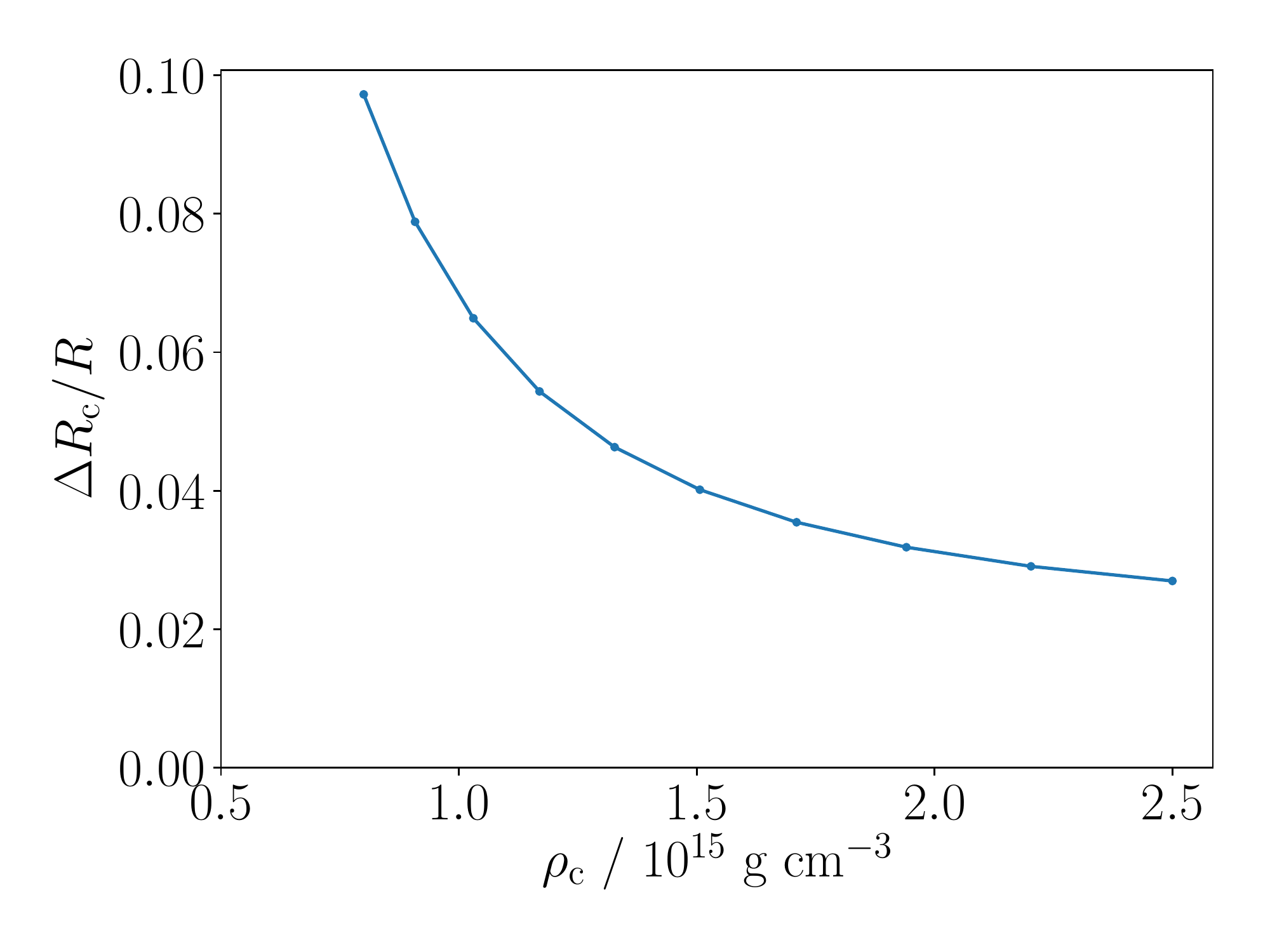}%
    \includegraphics[width=0.49\textwidth, keepaspectratio]{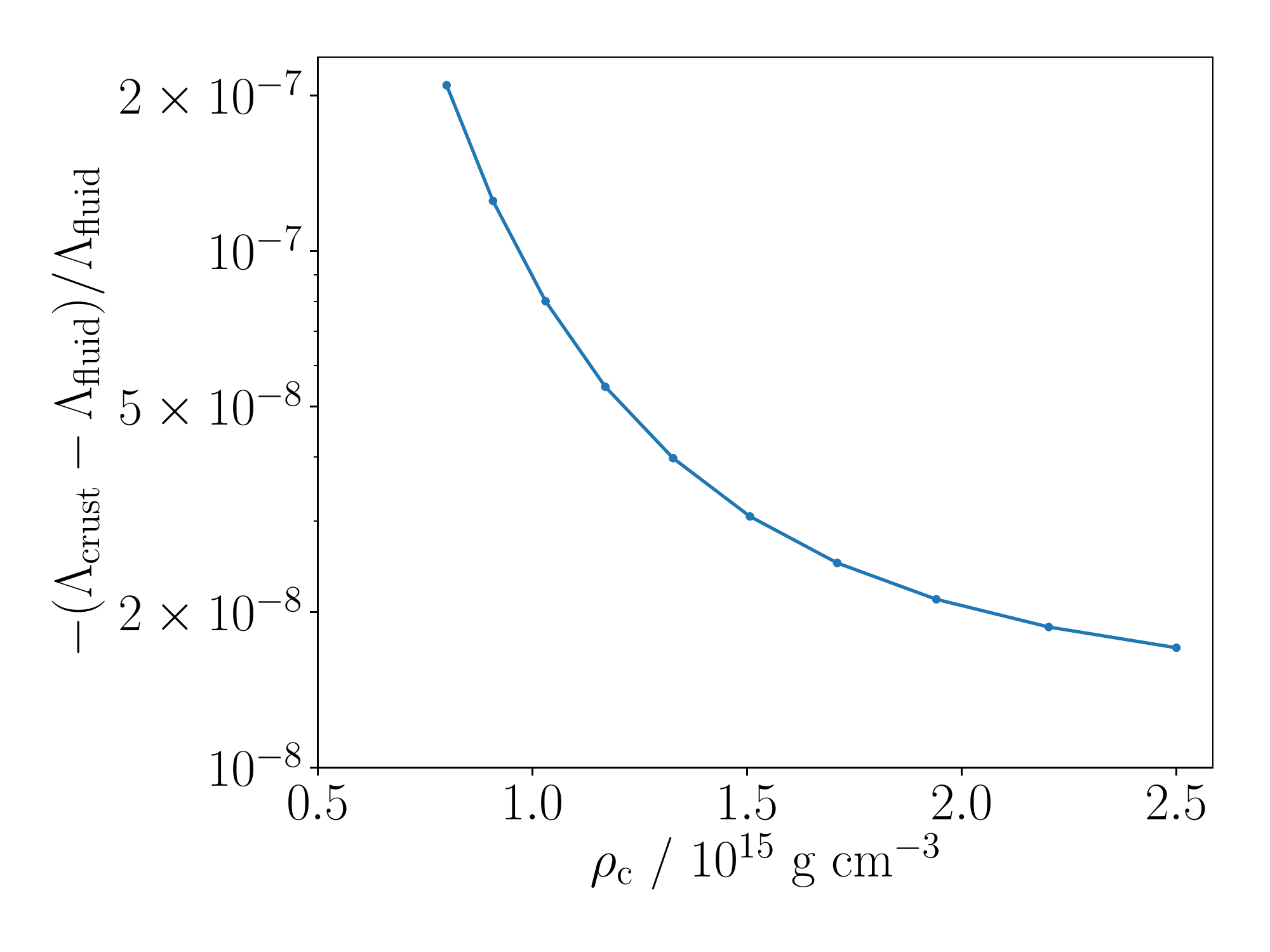}%
	\caption{\label{fig:ThicknessLove}The ratio of crustal thickness to stellar 
			 radius (left panel) 
             and the relative change in tidal deformability due to the 
             presence of a crust (right panel) as functions of central 
             density. As the central density approaches the core-crust 
			 transition (which occurs at 
			 $\rho_\text{base} = \SI{1.3e14}{\gram\per\centi\metre\cubed}$) 
             the crust occupies a much larger fraction of the star, so both 
             quantities become more significant.}
\end{figure}

For each stellar model, we compute the tidal deformability in the presence of an 
elastic crust, $\Lambda_\text{crust}$, as well as when a crust is not present, 
$\Lambda_\text{fluid}$, for comparison. We also calculate the thickness of the 
crust, $\Delta R_\text{c}$. We show these quantities in 
Fig.~\ref{fig:ThicknessLove} against the central density. In agreement with 
\cite{2011PhRvD..84j3006P}, we find that the inclusion of an elastic crust has 
an almost negligible impact on the tidal deformability -- the correction is the 
largest for the least compact stars at around two parts in $\num{e7}$. This is 
because, as the compactness decreases, the crust takes up a much larger fraction 
of the star. Moreover, as one would expect, the crust works to resist the star's 
deformation which is why the tidal deformabilities computed with a crust are 
smaller.

\begin{figure}
    \includegraphics[width=0.7\textwidth, keepaspectratio]{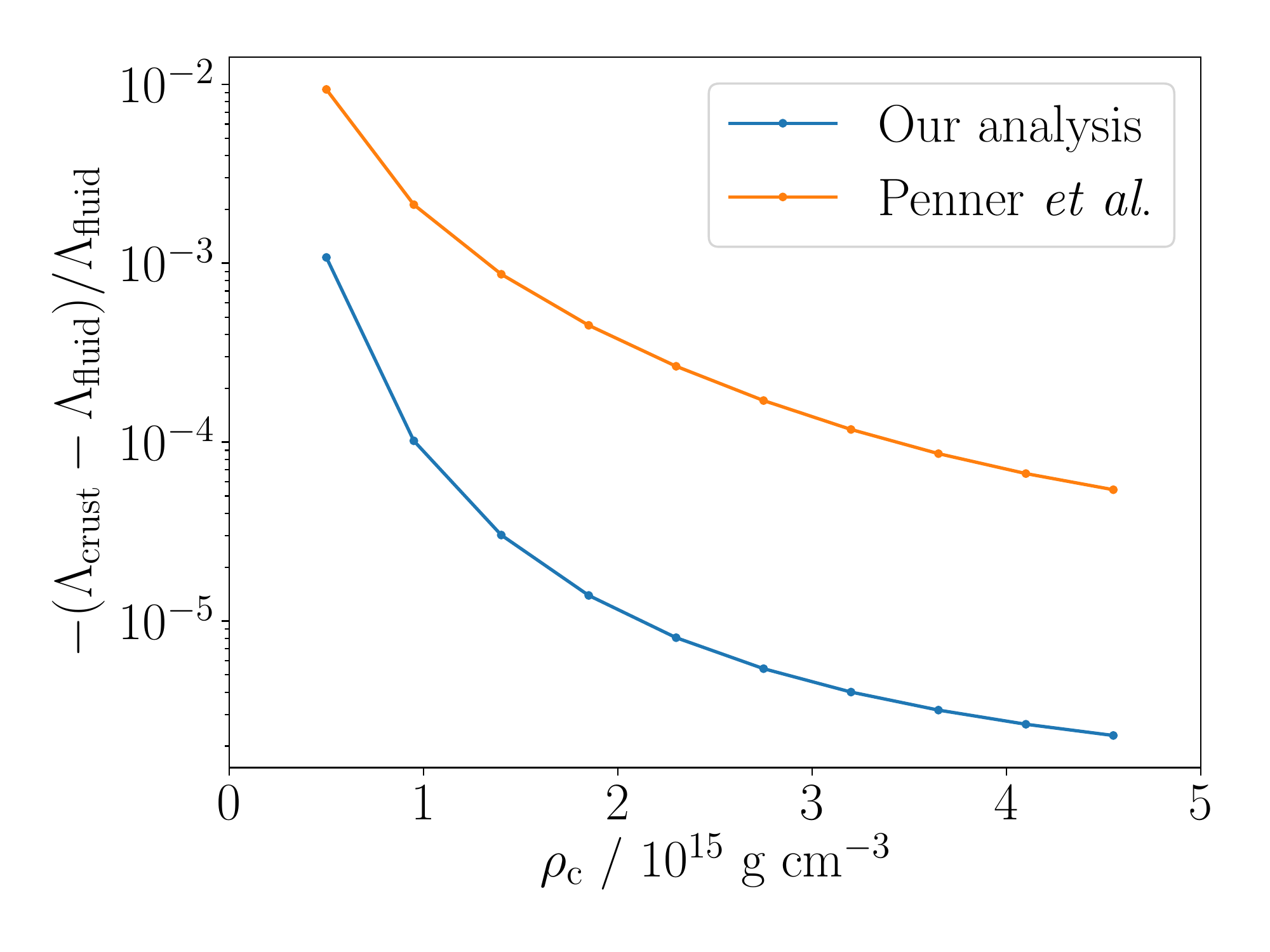}%
	\caption{\label{fig:LoveCompare}The relative change in the tidal 
			 deformability due to the presence of a crust against the central 
			 density for a polytropic equation of state with a linear shear 
			 modulus. We compare our results (blue) with those of 
			 \cite{2011PhRvD..84j3006P} (orange).}
\end{figure}

To facilitate direct comparison with \cite{2011PhRvD..84j3006P} we also 
integrated the perturbation equations with a polytropic equation of state and a 
shear modulus that scales linearly with the pressure. We used the same 
parameters as \cite{2011PhRvD..84j3006P} and moved the core-crust transition to 
$\rho_\text{base} = \SI{2e14}{\gram\per\centi\metre\cubed}$ and the crust-ocean 
transition to $\rho_\text{top} = \SI{e7}{\gram\per\centi\metre\cubed}$. The 
result is shown in Fig.~\ref{fig:LoveCompare}. In our calculation, we find that 
the tidal deformability is approximately an order of magnitude less sensitive to 
the inclusion of an elastic crust than reported by \cite{2011PhRvD..84j3006P}. 
This quantifies the effect of the error from (\ref{eq:PEErr}). (It is 
interesting to note that the crust has a more significant effect in this simple 
model as compared to the results from the realistic equation of state.)

Furthermore, we note that our results are in stark contrast with those of 
\cite{2019PhRvD.100d4056B} who find that the crust can make corrections to the 
tidal deformability of the order of $\sim 1 \%$. The reason for this 
disagreement is twofold. Firstly, \cite{2019PhRvD.100d4056B} write down 
expressions for the components of the displacement vector in the fluid and, 
thus, supposedly compute them in the fluid. This enables them to treat the 
system of coupled ordinary differential equations as an initial-value problem 
for the entire star and they use the continuity of the traction in order to 
match the fluid and elastic regions. As we noted in Sec.~\ref{sec:Fluid}, due to 
the static nature of the problem, extracting equations for the components of the 
displacement vector in the fluid is impossible. Additionally, by computing the 
perturbations as an initial-value problem means that one does not have the 
necessary freedom to enforce the traction conditions to be satisfied at the 
top of the crust, since the boundary conditions at the centre and the continuity 
conditions at the core-crust interface are sufficient to carry out the 
integrations. The second reason is due to the fact that 
\cite{2019PhRvD.100d4056B} do not have an outer fluid ocean in their stellar 
model, but instead have an exposed crust. In such a model, $H_0'$ is 
discontinuous and, therefore, one cannot use (\ref{eq:k_2}) as they do in order 
to compute the Love number. However, we note that the shear modulus at the top 
of the crust is expected to be small and so the discontinuity in $H_0'$ will be 
small. The difference in these results is important. If one assumes that 
third-generation gravitational-wave detectors will be able to constrain 
$\Lambda$ to within a few percent \cite{2020JCAP...03..050M}, then our results 
show that the effect of the crust will not be measurable, which is at odds with 
the results of \cite{2019PhRvD.100d4056B}.

\subsection{Crustal failure}

The formalism above allows us to calculate the interior structure of a 
neutron star with an elastic crust that is experiencing static, even-parity 
perturbations. We can apply this formalism to determine when and where the 
crust will begin to fracture during a binary neutron star inspiral, as was done 
in \cite{2012ApJ...749L..36P}. In contrast to the computation of the tidal Love 
number, the amplitude of the perturbations is important for this calculation. 
Therefore, we must normalise our perturbations by matching the interior solution 
to the exterior at the surface and, thus, constrain the amplitude.

We consider a binary separated by distance $d$ where the companion star is of 
mass $\Mcomp$. We assume $d \gg r$, as is appropriate in the adiabatic 
regime, and work in the Newtonian limit for the normalisation. By Kepler's third 
law, the angular frequency of the binary $\Omega$ is given by 
\begin{equation}
	\Omega^2 = \frac{M + \Mcomp}{d^3}.
\label{eq:KeplersThirdLaw}
\end{equation}
This is related to the orbital frequency of the binary $f_\text{orbit}$ by 
$\Omega = 2 \pi f_\text{orbit}$. First, let us estimate the gravitational-wave 
frequency at merger for an equal-mass binary, $M = \Mcomp$. We assume 
that the point of merger corresponds to when the two stars touch, $d = 2 R$, and 
since gravitational waves radiate at twice the orbital frequency, 
$f_\text{GW} = 2 f_\text{orbit}$, we find 
\begin{equation}
	f_\text{GW}^\text{merger} = \frac{1}{2 \pi} \sqrt{\frac{M}{R^3}} 
	\approx \num{2170} \, \left( \frac{M}{\SI{1.4}{\solarMass}} \right)^{1/2} 
	\left( \frac{R}{\SI{10}{\kilo\metre}} \right)^{-3/2} \, \si{\hertz}. 
\label{eq:MergerFrequency}
\end{equation}

Expanding around $r = 0$ one can show that the external field due to the 
presence of the companion is 
\begin{equation}
	\Phi_\text{ext}(x^i) = - \frac{\Mcomp}{d} 
	- \frac{\Mcomp}{d^2} r m_i n^i 
	- \frac{3}{2} \frac{\Mcomp}{d^3} r^2 
	\left( m_i m_j - \frac{1}{3} \delta_{i j} \right) n^i n^j, 
\label{eq:ExternalField}
\end{equation}
where $m^i$ is the unit vector that points from the centre of the star to 
the centre of the companion. The tidal piece can be expressed using the 
$l = 2$, $m = 0$ spherical harmonic, 
\begin{equation}
	\Phi_\text{tidal}(x^i) 
	= - \sqrt{\frac{4 \pi}{5}} \frac{\Mcomp}{d^3} r^2 Y_{2 0}.
\end{equation}
This means that the binary is orientated such that $\theta = 0$ points in the 
direction of $m^i$. The tidal multipole in the Newtonian limit is given by, 
using (\ref{eq:ExternalField}),
\begin{equation}
	\mathcal{E}_{i j} 
	= \frac{\partial^2 \Phi_\text{ext}}{\partial x^i \partial x^j} 
	= - 3 \frac{\Mcomp}{d^3} 
	\left( m_i m_j - \frac{1}{3} \delta_{i j} \right).
\end{equation}
Using the decomposition of (\ref{eq:Decomposition}) one can show that the 
non-vanishing $\mathcal{E}_{2 m}$ is 
\begin{equation}
	\mathcal{E}_{2 0} = - 2 \sqrt{\frac{4 \pi}{5}} \frac{\Mcomp}{d^3},
\end{equation}
and, therefore, by (\ref{eq:c_2a}) we find 
\begin{equation}
	c_2 = \frac{2}{3} \sqrt{\frac{4 \pi}{5}} \frac{M^2 \Mcomp}{d^3} 
	= \frac{2 \pi^2}{3} \sqrt{\frac{4 \pi}{5}} 
	\frac{M^2 \Mcomp}{M + \Mcomp} f_\text{GW}^2.
\label{eq:c_2b}
\end{equation}
Here we have chosen to parametrise the point in the inspiral by the 
gravitational-wave frequency over the separation by using 
(\ref{eq:KeplersThirdLaw}). Eqs.~(\ref{eq:Ratio}) and (\ref{eq:c_2b}) provide 
the necessary information to normalise the perturbations to a binary that is 
emitting gravitational waves with frequency $f_\text{GW}$.

We use the von Mises criterion to determine when the crust begins to break. 
In the formalism of \cite{2019CQGra..36j5004A}, one can calculate the von Mises 
strain for a neutron star, where the unperturbed configuration is unstrained, 
through 
\begin{equation}
    \Theta = \sqrt{\frac{3}{2} \Delta s_{a b} \Delta s^{a b}}.
\end{equation}
The crust fractures when the von Mises strain reaches the threshold yield point 
$\Theta \geq \Theta^\text{break}$. Using the definition of the traction 
variables (\ref{eq:TractionVariables}) with (\ref{eq:Strain}) and specialising 
to $l = 2$, $m = 0$ perturbations, one finds 
\begin{equation}
    \Theta^2 = \frac{45}{256 \pi} \frac{1}{r^4} 
    \left[ (3 \cos^2 \theta - 1)^2 \left( \frac{T_1}{\muc} \right)^2 
    + 12 e^{-\lambda} \sin^2(2 \theta) \left( \frac{T_2}{\muc} \right)^2 
    + 48 \sin^4 \theta V^2 \right].
\label{eq:vonMises}
\end{equation}
The advantage of using the von Mises strain is that it is a function of 
position, and so we can identify where the crust is the weakest as well as when 
it breaks. Taking the breaking strain to be $\Theta^\text{break} = 0.1$ 
\cite{2009PhRvL.102s1102H}, we can calculate when the crust will break, at each 
point, by imposing that the strain in (\ref{eq:vonMises}) is equal to 
$\Theta^\text{break}$ to normalise the perturbations and then determining the 
gravitational-wave frequency $f_\text{GW}^\text{break}$ which corresponds to 
that amplitude using (\ref{eq:Ratio}) and (\ref{eq:c_2b}).

\begin{figure}
	\includegraphics[width=0.9\textwidth]{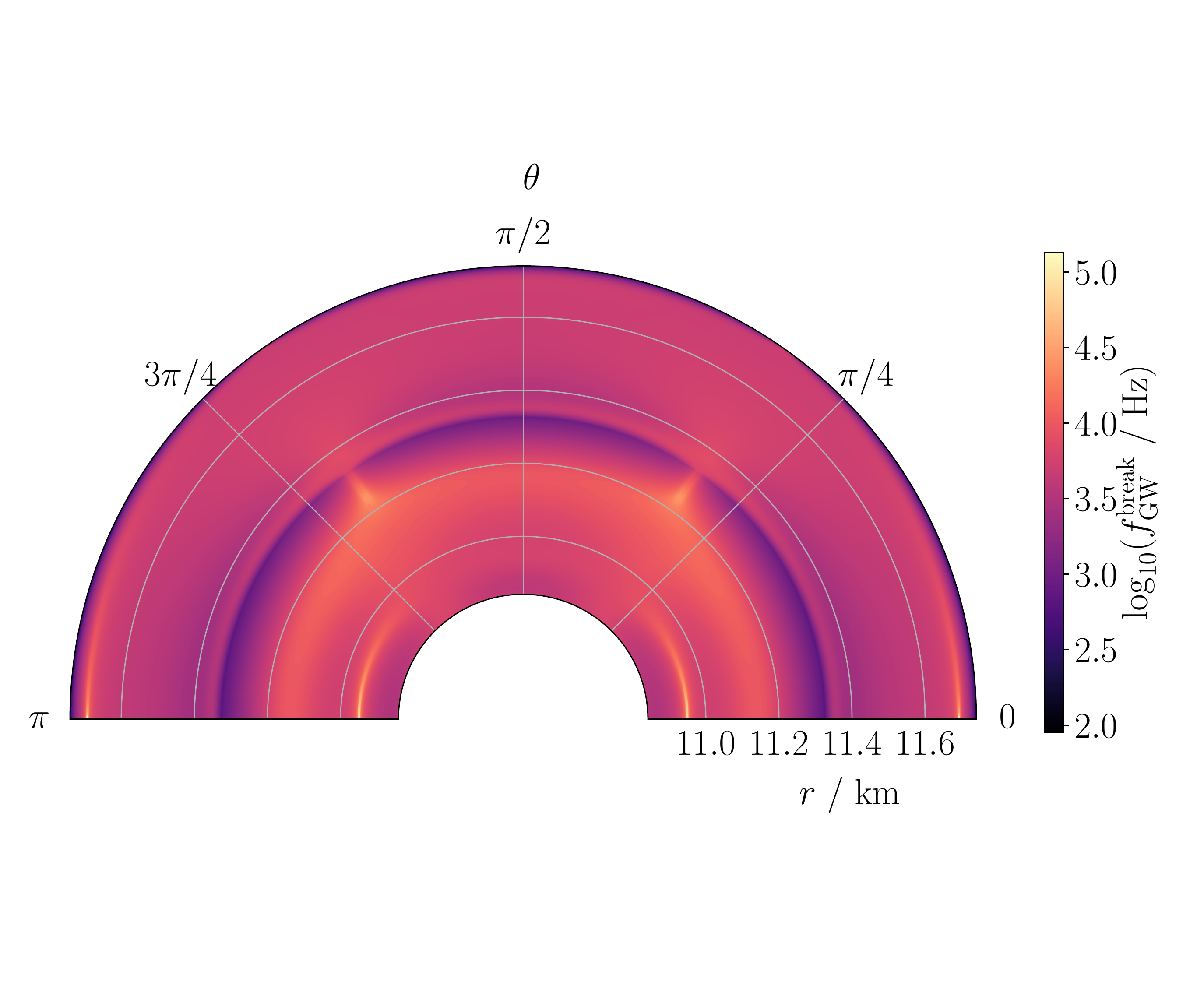}%
	\caption{\label{fig:BreakingFrequency}The gravitational-wave frequency at 
			 failure across the elastic crust. We can see that the majority of 
			 the crust will not fail before merger.}
\end{figure}

\begin{figure}
	\includegraphics[width=0.9\textwidth]{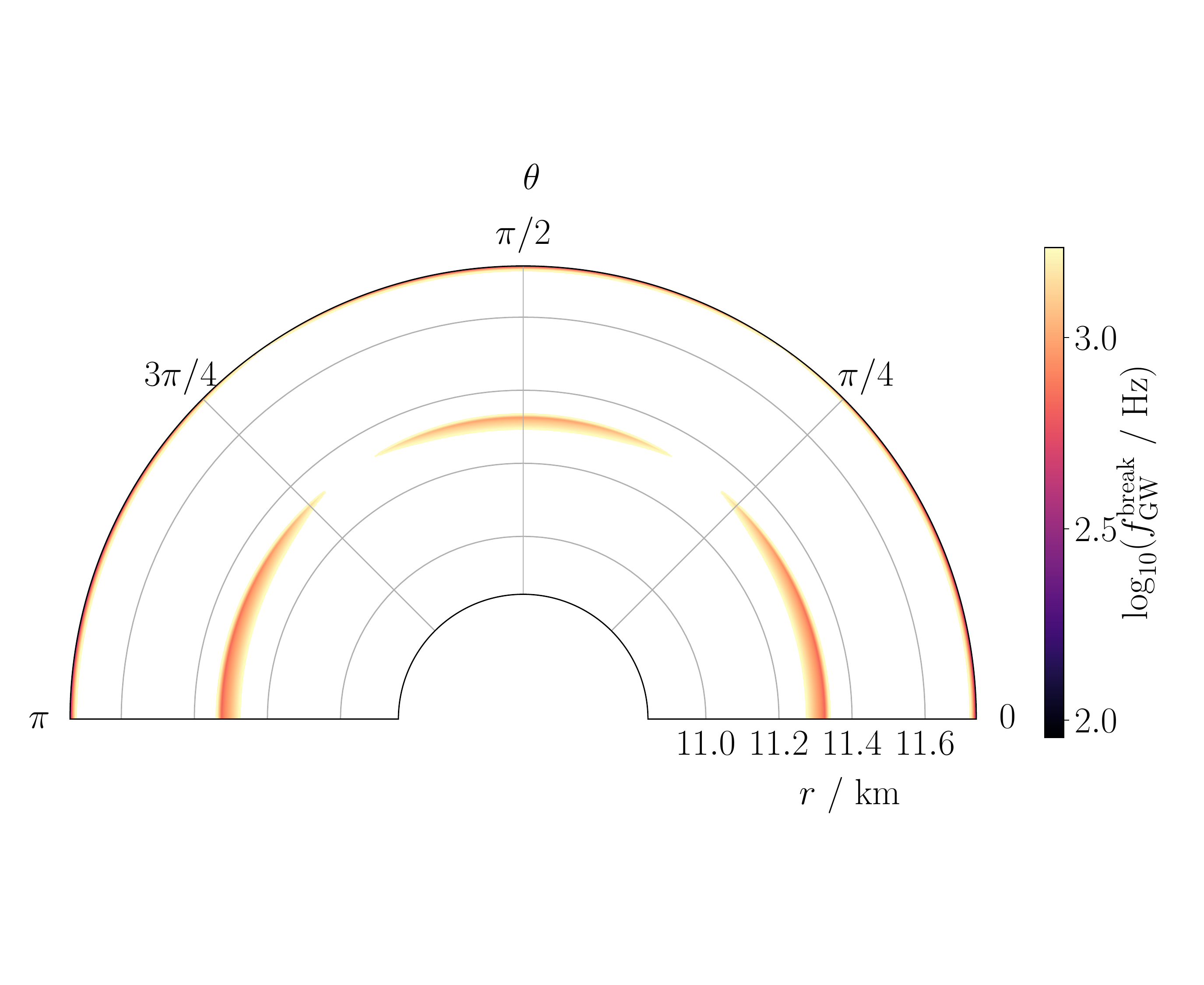}%
	\caption{\label{fig:BreakingFrequencySaturate}The gravitational-wave 
			 frequency at failure for the locations in the elastic crust that 
			 yield before merger. This shows that the stars will merge with the 
			 crust largely intact.}
\end{figure}

\begin{figure}
	\includegraphics[width=0.7\textwidth]{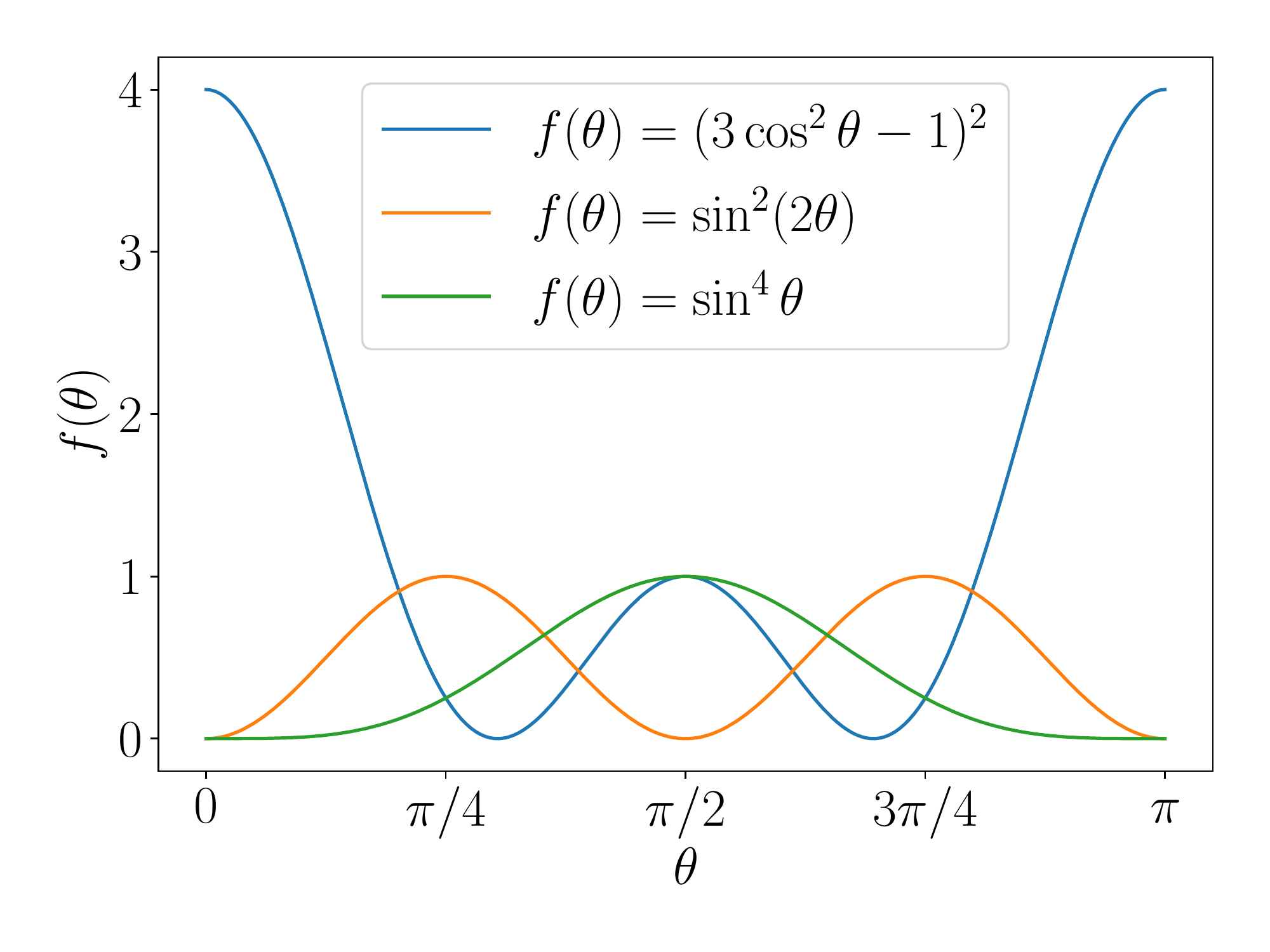}%
	\caption{\label{fig:AngularBasis}The angular basis of the von Mises strain 
	for $l = 2$, $m = 0$ perturbations.}
\end{figure}

\begin{figure}
	\includegraphics[width=0.7\textwidth]{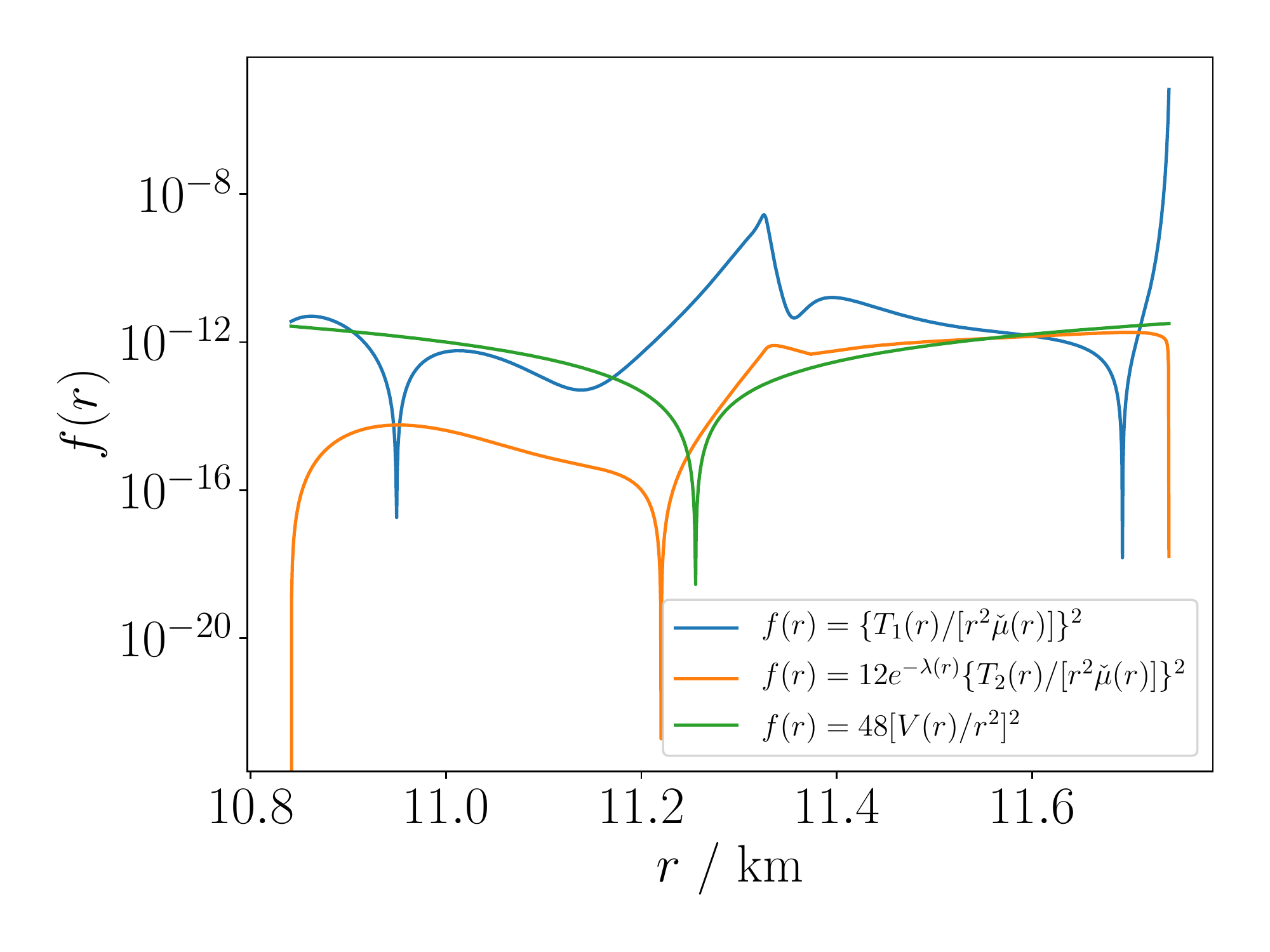}%
	\caption{\label{fig:RadialBasis}The radial dependence of the radial 
	and tangential traction variables and the tangential displacement function 
	normalised 	to a binary radiating gravitational waves with 
	$f_\text{GW} = \SI{10}{\hertz}$. At neutron drip, around 
	$r = \SI{11.3}{\kilo\metre}$, we find that the tangential functions combine 
	to give local minima for most values of $\theta$. Note that the vertical 
	axis is in logrithmic scale -- the cusps correspond to when the functions 
	cross zero and change sign.}
\end{figure}

As an illustration, we use the same equation of state as in Sec.~\ref{sec:k_2}. 
We assume the binary is equal mass with 
$M = \Mcomp = \SI{1.4}{\solarMass}$, for which we obtain a star with 
radius $R = \SI{11.74}{\kilo\metre}$. In Fig.~\ref{fig:BreakingFrequency} we 
show the gravitational-wave frequency when the crust breaks at each point. 
Fig.~\ref{fig:BreakingFrequencySaturate} focuses on the regions of the star 
that break before merger. There is a clear phase transition at neutron drip 
(around $r = \SI{11.3}{\kilo\metre}$), where the inner crust is, on average, 
stronger than the outer crust. The crust is notably strong at neutron drip 
close to $\theta \approx \pi / 4$ and $3 \pi / 4$. The reason for this is, as 
the star becomes more oblate, the parts closest to the poles and equator are 
stretched the most. The region that stretches the least is in between these two 
regions at $\theta \approx \pi / 4$ and $3 \pi / 4$. This effect can be seen in 
Figs.~\ref{fig:AngularBasis} and \ref{fig:RadialBasis} where the tangential 
functions $T_2 / \muc$ and $V$ combine to give a local minimum in the von Mises 
strain at these angles and, thus, a local maximum in the breaking frequency. The 
maxima along the equator, $\theta = 0$ and $\pi$, are where 
the magnitude of the radial traction function $T_1 / \muc$ reaches a local 
minimum (as shown in Fig.~\ref{fig:RadialBasis}).

We note that, as compared to typical merger frequencies 
(\ref{eq:MergerFrequency}), our results suggest that the vast majority of the 
crust will not fracture before merger. In fact, the crust will only fail at 
neutron drip and in the very outermost part of the outer crust before 
coalescence. This is in contrast to the results of \cite{2012ApJ...749L..36P}, 
who obtain significantly lower breaking frequencies throughout the crust. This 
is likely related to the errors in the analysis of \cite{2011PhRvD..84j3006P}, 
that we have pointed out above. Since the crust will be mostly intact by the 
point of merger, this suggests that there will not be a significant amount of 
strain energy released available for an associated electromagnetic signal.

\section{\label{sec:Conclusions}Conclusions}

With the advent of gravitational-wave detections of binary neutron star mergers, 
we have a promising new method of constraining the equation of state of nuclear 
matter. The gravitational waveforms from these events are sensitive to tidal 
effects in the binaries which carry model-independent information on the 
equation of state.

In this paper, we have explored the impact of an elastic crust on tidal 
deformations of neutron stars. We have presented a formalism which enables one 
to compute static, even-parity perturbations of a neutron star with an elastic 
component. This was necessary to resolve discrepancies between previous studies 
\cite{2011PhRvD..84j3006P, 2019PhRvD.100d4056B}. There are mistakes in the 
crustal perturbation equations presented by \cite{2011PhRvD..84j3006P}, in 
particular arising from the analogous equation to (\ref{eq:PEErr}). This meant 
they marginally overestimated the impact of a crust on tidal deformations and 
consequently this affected their analysis on when the crust will break in a 
binary inspiral \cite{2012ApJ...749L..36P}. Meanwhile, the work of 
\cite{2019PhRvD.100d4056B} calculates the static displacement vector in the 
fluid regions of the star. However, such a calculation should not be possible 
due to the static nature of the problem. This means that they cannot correctly 
impose continuity of the traction at the top of the crust. Moreover, 
\cite{2019PhRvD.100d4056B} do not correctly calculate the tidal Love number 
for the assumed stellar model with an exposed crust.

We have applied our formalism to the computation of static, quadrupolar 
perturbations of a neutron star sourced by an external tidal field. We 
calculated the quadrupolar perturbations for a realistic equation of state that 
includes an elastic crust. We have shown that the inclusion of an elastic crust 
has a very small effect on the tidal deformability of a star, in the range of 
$\sim \numrange{e-8}{e-7}$ for realistic models -- even smaller than what one 
would calculate using simplistic equations of state. We found that our results 
are an order of magnitude smaller than what was reported by 
\cite{2011PhRvD..84j3006P} and significantly smaller than the results of 
\cite{2019PhRvD.100d4056B}. This means the impact of a crust on binary neutron 
star mergers is not expected to be detectable for current and next-generation 
gravitational-wave detectors.

We used our integrations to calculate when and where the crust would fail during 
a binary inspiral with component masses 
$M = \Mcomp = \SI{1.4}{\solarMass}$. We found that the crust is much stronger 
than estimated in previous work \cite{2012ApJ...749L..36P}. The majority of the 
crust will not fail before the two neutron stars merge. Only the small regions 
close to neutron drip and the outer layers of the crust will fracture before 
merger.

\begin{acknowledgments}
NA gratefully acknowledges financial support from STFC via Grant No. 
ST/R00045X/1. JPP is thankful for the partial support by the Polish National 
Science Centre via Grant No. 2016/22/E/ST9/00037 and FAPESP via Grants No. 
2015/04174-9 and No. 2017/21384-2.
\end{acknowledgments}

% Specify following sections are appendices. Use \appendix* if there
% only one appendix.
\appendix
\section{\label{app:Interface}Calculating the interface conditions}

Since we consider a star with multiple layers that have phase transitions, we 
must address how the perturbation functions behave across an interface. We 
calculate the interface conditions using the geometrical approach explained in 
\cite{2002PhRvD..66j4002A}.

Let us begin by considering the level surfaces of a scalar quantity $A$. We 
assume the level surfaces to be timelike and, therefore, have the normal, 
\begin{equation}
	\mathcal{N}^a = \frac{\partial^a A}{\sqrt{\partial^b A \partial_b A}},
\end{equation}
where $\mathcal{N}^a \mathcal{N}_a = 1$ is true by construction. The first 
fundamental form (also known as the intrinsic curvature or induced three-metric) 
of these level surfaces is 
\begin{equation}
	\gamma_{a b} 
	= P_a^{\hphantom{a} c} P_b^{\hphantom{b} d} g_{c d}, 
\end{equation}
where the projection operator along the level surfaces is given by 
\begin{equation}
	P_a^{\hphantom{a} b} = \delta_a^{\hphantom{a} b} 
	- \mathcal{N}_a \mathcal{N}^b.
\end{equation}
The second fundamental form (also known as the extrinsic curvature) of the level 
surfaces is defined as 
\begin{equation}
	K_{a b} 
	= - P_a^{\hphantom{a} c} P_b^{\hphantom{b} d} 
	\nabla_{(c} \mathcal{N}_{d)}.
\end{equation}

Let us specialise and consider the useful decomposition of our scalar quantity 
of the form, 
\begin{equation}
    	A(t, r, \theta, \phi) = A_0(r) + \delta A(t, r, \theta, \phi).
\end{equation}
Using this decomposition we obtain the following components for the normal: 
\begin{subequations}
\begin{align}
	\mathcal{N}^t &= - e^{- \nu + \lambda/2} 
	\frac{\partial_t \delta A}{A_0'} + e^{- \nu - \lambda/2} h_{t r}, \\
	\mathcal{N}^r &= e^{-\lambda/2} 
	\left( 1 - \frac{1}{2} e^{-\lambda} h_{r r} \right), \\
	\mathcal{N}^\theta &= \frac{e^{\lambda/2}}{r^2} 
	\frac{\partial_\theta \delta A}{A_0'}, \\
	\mathcal{N}^\phi &= \frac{e^{\lambda/2}}{r^2 \sin^2 \theta} 
	\frac{\partial_\phi \delta A}{A_0'}.
\end{align}
\end{subequations}
The level surfaces of $A$, thus, have the following non-zero components of the 
first fundamental form: 
\begin{subequations}
\begin{align}
    \gamma_{t t} &= - e^\nu + h_{t t}, \\
    \gamma_{t r} &= h_{t r} - e^\lambda \frac{\partial_t \delta A}{A_0'}, \\
   	 \gamma_{r \theta} &= - e^\lambda \frac{\partial_\theta \delta A}{A_0'}, \\
    \gamma_{r \phi} &= - e^\lambda \frac{\partial_\phi \delta A}{A_0'}, \\
    \gamma_{\theta \theta} &= r^2 + h_{\theta \theta}, \\
    \gamma_{\phi \phi} &= r^2 \sin^2 \theta + h_{\phi \phi}.
\end{align}
\end{subequations}
The non-trivial components of the second fundamental form are 
\begin{subequations}
\begin{align}
	K_{t t} &= \frac{\nu'}{2} e^{\nu - \lambda/2} - e^{\lambda/2} 
	\frac{\partial_t^2 \delta A}{A_0'} + e^{-\lambda/2} \partial_t h_{t r} 
	- \frac{1}{2} e^{-\lambda/2} h_{t t}' 
	- \frac{\nu'}{4} e^{\nu - 3\lambda/2} h_{r r}, \\
	K_{t r} &= \frac{\nu'}{2} \left( e^{\lambda/2} 
	\frac{\partial_t \delta A}{A_0'} - e^{-\lambda/2} h_{t r} \right), \\
	K_{t \theta} &= - e^{\lambda/2} 
	\left( \frac{\partial_t \partial_\theta \delta A}{A_0'} 
	- \frac{1}{2} e^{-\lambda} \partial_\theta h_{t r} \right), \\
	K_{t \phi} &= - e^{\lambda/2} 
	\left( \frac{\partial_t \partial_\phi \delta A}{A_0'} 
	- \frac{1}{2} e^{-\lambda} \partial_\phi h_{t r} \right), \\
	K_{r \theta} &= \frac{e^{\lambda/2}}{r} 
	\frac{\partial_\theta \delta A}{A_0'} \\
	K_{r \phi} &= \frac{e^{\lambda/2}}{r} 
	\frac{\partial_\phi \delta A}{A_0'}, \\
	K_{\theta \theta} &= - e^{-\lambda/2} r - e^{\lambda/2} 
	\frac{\partial_\theta^2 \delta A}{A_0'} 
	- \frac{1}{2} e^{-\lambda/2} ( h_{\theta \theta}' 
	- e^{-\lambda} r h_{r r} ), \\
	K_{\theta \phi} &= - \frac{e^{-\lambda/2}}{A_0'} 
	\left( \partial_\theta \partial_\phi \delta A 
	- \cot\theta \partial_\phi \delta A \right), \\
	K_{\phi \phi} &= - e^{-\lambda/2} r \sin^2 \theta - e^{\lambda/2} 
	\frac{\partial_\phi^2 \delta A}{A_0'} 
	- e^{\lambda/2} \sin\theta \cos\theta \frac{\partial_\theta \delta A}{A_0'} 
	- \frac{1}{2} e^{-\lambda/2} ( h_{\phi \phi}' 
	- e^{-\lambda} r \sin^2 \theta h_{r r} )
\end{align}
\end{subequations}
Both the first and second fundamental forms must be continuous across an 
interface (in the absence of surface degrees of freedom).

As was done by \cite{1990MNRAS.245...82F}, we will consider the level surfaces 
of the radial shell, so we assign $A_0 = r$ and $\delta A = \xi^r$. We use the 
perturbed metric for even-parity perturbations (\ref{eq:PerturbedMetric}). 
Because of how we set up the problem by assuming the background star is in a 
relaxed state, we know that the background quantities will all be continuous 
across an interface. We further assume that there is no discontinuity in the 
density or pressure. The first fundamental form with components $\gamma_{t t}$, 
$\gamma_{t r}$, $\gamma_{\theta \theta}$ and $\gamma_{r \theta}$ show 
\begin{equation}
    [H_0]_r = 0, \quad
    [H_1]_r = 0, \quad
   	[K]_r = 0, \quad 
    [\delta A / A_0']_r = 0, 
\end{equation}
where we have introduced the notation 
$[f]_r = \lim_{\epsilon \rightarrow 0} [f(r + \epsilon) - f(r - \epsilon)]$ to 
describe the continuity of a function $f(r)$ at a point $r$. The angular part of 
$\delta A$ is decomposed using spherical harmonics. For the problem we 
are analysing, $H_1$ simply vanishes. The generic condition above translates to 
$[\xi^r]_r = 0$, which is equivalent to 
\begin{equation}
    	[W]_r = 0.
\end{equation}
This condition is equivalent to saying there must not be a gap in the perturbed 
material.

We have exhausted the information we can learn from continuity of the first 
fundamental form. We also notice that there is no additional information 
to be learned from the components $K_{t r}$, $K_{t \theta}$, $K_{t \phi}$,  
$K_{r \theta}$ and $K_{r \phi}$ of the second fundamental form; only components 
$K_{t t}$ and $K_{\theta \theta}$ provide more interface conditions. Continuity 
of $K_{t t}$ implies 
\begin{equation}
    \left[ h_{t t}' + \frac{\nu'}{2} e^{\nu - \lambda} h_{r r} \right]_r = 0.
\end{equation}
This gives 
\begin{equation}
    [H_0']_r = - \frac{\nu'}{2} [H_2]_r.
\label{eq:H_0primea}
\end{equation}
Similarly, we can infer from $K_{\theta \theta}$ 
\begin{equation}
	[h_{\theta \theta}' - e^{-\lambda} r h_{r r}]_r 
	= 0 \quad \Rightarrow \quad [K']_r = \frac{1}{r} [H_2]_r.
\label{eq:Kprime}
\end{equation}
We can combine (\ref{eq:H_0primea}) and (\ref{eq:Kprime}) to obtain 
\begin{equation}
    [K' - H_0']_r = \frac{1}{2 r} (2 + r\nu') [H_2]_r.
\end{equation}

Now, we need some information from the perturbed Einstein equations. The above 
expression can be further used along with (\ref{eq:PEErthetaa}) to provide 
\begin{equation}
	\frac{16 \pi}{r} (2 + r \nu') [\muc V]_r - \frac{16 \pi}{r} [T_2]_r 
	= \frac{1}{2 r} (2 + r \nu') [H_2]_r.
\label{eq:PEErtheta2Continuity}
\end{equation}
Using continuity of $H_0$ and (\ref{eq:PEEdifference}) we find 
\begin{equation}
    [H_2]_r = 32 \pi [\muc V]_r.
\label{eq:PEEdifferenceContinuity}
\end{equation}
This condition states that we should expect a discontinuity in $H_2$ for two 
reasons: (i) the shear modulus vanishes in the fluid and has a finite value in 
the crust, and (ii) there is no reason that the tangential displacement function 
$V$ need be continuous. This further implies through (\ref{eq:H_0primea}) that 
$H_0'$ will be discontinuous, 
\begin{equation}
	[H_0']_r = - 16 \pi \nu' [\muc V]_r.
\label{eq:H_0primeb}
\end{equation}
Eqs.~(\ref{eq:PEErtheta2Continuity}) and 
(\ref{eq:PEEdifferenceContinuity}) imply continuity of the tangential traction 
variable, 
\begin{equation}
    [T_2]_r = 0.
\label{eq:T_2Continuity}
\end{equation}

Finally, we use (\ref{eq:AlgebraicRelationa}), along with the continuity 
condition (\ref{eq:H_0primeb}), to obtain 
\begin{equation}
    [T_1 + r^2 \delta p]_r = 0.
\end{equation}
Since the radial displacement function $W$ is continuous, we can write this in a 
more general form, 
\begin{equation}
    	[T_1 + r^2 \Delta p]_r = 0.
\label{eq:T_1Continuity}
\end{equation}
Eqs.~(\ref{eq:T_2Continuity}) and (\ref{eq:T_1Continuity}) simply mean that the 
radial and tangential stresses are continuous across a fluid-elastic interface. 
These interface conditions are necessary when considering how the functions 
behave across a fluid-elastic boundary and enable one to carry out the 
integration in the crust.

\section{\label{app:Numerical}Numerical scheme}

Our approach to solving the interior perturbation equations is similar to as 
described in \cite{2008PhRvD..78h3008L, 2015PhRvD..92f3009K}. We 
divide our star into three layers: (i) a fluid core from $R_0 = 0$ to $R_1$, 
(ii) an elastic crust from $R_1$ to $R_2$, and (iii) a fluid ocean from 
$R_2$ to $R_3 = R$. We express the system of ordinary differential equations 
for a given layer $i$ in the form 
\begin{equation}
    \frac{d \mathbf{Y}^{(i)}}{d r} = \mathbf{Q}^{(i)} \cdot \mathbf{Y}^{(i)}, 
    \ \text{for} \ r \in [R_{i - 1}, R_i],
\label{eq:ODE}
\end{equation}
where $\mathbf{Y}^{(i)}(r) = [y_1(r), ..., y_{k_i}(r)]$ is an abstract 
$k_i$-dimensional vector field, $\mathbf{Q}^{(i)}(r)$ is a $k_i \times k_i$ 
matrix and $r = R_i$ denotes the end of layer $i$. As long as our differential 
equations are linear we are free to write the system in the above form.

Due to the linearity of the differential equations, we generate a set of $k_i$ 
linearly-independent solutions $\mathbf{Y}_j^{(i)}(r)$ for layer $i$ and obtain 
the general solution using a linear combination of these solutions, 
\begin{equation}
    \mathbf{Y}^{(i)}(r) = \sum_{j = 1}^{k_i} c_j^{(i)} \mathbf{Y}_j^{(i)}(r), 
\end{equation}
where the coefficients $c_j^{(i)}$ are constants to be determined from 
boundary and interface conditions. We generate these linearly-independent 
solutions by choosing linearly-independent start vectors 
$\mathbf{Y}_j^{(i)}(R_{i - 1})$ and integrating through the layer using 
(\ref{eq:ODE}) up to $r = R_i$. (Note that, in theory, there is no 
reason one could not do the reverse, integrating from $r = R_i$ to $R_{i - 1}$, 
should they wish.) \textit{A priori}, we do not have any 
additional information about layer $i$ and would na{\"i}vely integrate $k_i$ 
linearly-independent start vectors. However, we can reduce the computational 
effort by applying relevant boundary conditions. For example, should a variable 
vanish at an interface, one could simply set this variable to zero in the start 
vectors and reduce the number of necessary linearly-independent solutions by 
one.

The fluid regions of the star are governed by Eqs.~(\ref{eq:FluidODEs})
and so are fully described by the abstract two-dimensional vector field, 
\begin{equation}
    \mathbf{Y}^{(k)}(r) = [H_0'(r), H_0(r)],
\end{equation}
where $k = 1, 3$ denotes the core and ocean, respectively.
The elastic region of the star is more complex and 
requires more functions to describe its structure. Thus, we use the 
six-dimensional vector field, 
\begin{equation}
    \mathbf{Y}^{(2)}(r) = [H_0'(r), H_0(r), K(r), W(r), V(r), T_2(r)].
\end{equation}
At a fluid-elastic interface we know the variables $H_0$, $K$, $W$ and $T_2$ 
are continuous. We know the values of $H_0$ and $K$ from the calculation in the 
fluid core, and so we use their final values in the core to start our 
integration in the crust. Since the traction variables vanish in the fluid we 
can simplify the integrations in the elastic by demanding that $T_2 = 0$ at an 
interface. For each of the solutions we calculate the value for 
$H_0'$ at the base using (\ref{eq:H_0primeCondition}). These conditions mean 
that we must generate two linearly-independent solutions with the initial values 
for the unknown functions $W$ and $V$. Notice that because of the condition 
(\ref{eq:H_0primeCondition}) we could equivalently choose to generate solutions 
with $H_0'$ instead of $V$. At the top of the crust we demand that $T_2 = 0$ and 
$H_0'$ be equal to the expression calculated using (\ref{eq:H_0primeCondition}). 
We use these two constraints to solve for the coefficients of the general 
solution. At the top of the crust, we can straightforwardly continue the 
integration through the fluid ocean, since $H_0$ and $K$ are continuous. 
Although, $H_0'$ is discontinuous, we can calculate it exactly from $H_0$ and 
$K$ using (\ref{eq:KFluid}) and integrate to the surface. 

% Create the reference section using BibTeX:
\bibliography{bibliography}

\end{document}